\documentclass[11pt,preprint]{aastex}
\usepackage{color}

\newcommand{\piZ}{$\pi^0$}
\newcommand{\synTherm}{$\frac{\mathrm{synch}}{\mathrm{therm}}$}
\newcommand{\pionIC}{(pion/IC)$_\mathrm{1TeV}$}
\newcommand{\pionICfr}{$\frac{\mathrm{pion}}{\mathrm{IC}}|_\mathrm{1TeV}$}
\newcommand{\SumR}{$\sum{\rho^2V}$}
\newcommand{\IE}{instant equilibration}
\newcommand{\CH}{Coulomb heating}
\newcommand{\EffRel}{\epsilon_\mathrm{rel}}
\newcommand{\EffEsc}{\epsilon_\mathrm{esc}}
\newcommand{\Acut}{\alpha_\mathrm{cut}}
\newcommand{\SA}{semi-analytic}

\newcommand{\alf}{Alfv\'en}

\newcommand{\gamrays}{$\gamma$-rays}
\newcommand{\NT}{non-thermal}
\newcommand{\NL}{nonlinear}
\newcommand{\muG}{$\mu$G}
\newcommand{\epRel}{(e/p)_{\mathrm{rel}}}
\newcommand{\SC}{self-consistent}
\newcommand{\SCly}{self-consistently}

\newcommand{\etamfp}{\eta_\mathrm{mfp}}
\newcommand{\CD}{contact discontinuity}

\newcommand{\pmax}{p_\mathrm{max}}

\newcommand{\Pmax}{p_\mathrm{max}}
\newcommand{\TP}{test-particle}

\newcommand{\RFS}{R_\mathrm{FS}}

\newcommand{\rg}{r_g}

\newcommand\Rtot{r_\mathrm{tot}}

\newcommand{\xx}[1]{\times 10^{#1}}

\newcommand{\rel}{relativistic}

\newcommand{\nonrel}{non-relativistic}
\newcommand{\NEI}{non-equilibrium ionization}

\newcommand{\Syn}{Synchrotron}
\newcommand{\syn}{synchrotron}
\newcommand{\synch}{synchrotron}
\newcommand{\brem}{bremsstrahlung}

\newcommand{\IC}{inverse-Compton}
\newcommand{\pion}{pion-decay}
\newcommand\tSNR{t_\mathrm{SNR}}
\newcommand\dSNR{D_\mathrm{SNR}}
\newcommand\EnSN{E_\mathrm{SN}}
\newcommand\Mej{M_\mathrm{ej}}

\newcommand\Inj{\chi_\mathrm{inj}}

\newcommand{\pcc}{cm$^{-3}$}

\newcommand\Msun{\mathrm{M}_{\odot}}

\newcommand{\denISM}{n_\mathrm{H}}
\newcommand{\denH}{n_\mathrm{H}}

\newcount\listnorom
\newcommand\listromanDE{\global\advance \listnorom by 1
{\lowercase\expandafter{(\romannumeral\listnorom)}\ }}
\newcommand\newlistroman{\listnorom=0}
\newcount\Inum
\newcount\IInum
\newcount\IIInum
\newcount\IVnum
\def\I{\global\multiply\IInum by 0 \global\multiply\IIInum by 0
            \global\multiply\IVnum by 0 \global\advance \Inum by 1
            {\the\Inum. }}
\def\II{\global\multiply\IIInum by 0\global\multiply\IVnum by 0
       \global\advance \IInum by 1 {\the\Inum.\the\IInum. }}
\def\III{\global\multiply\IVnum by 0\global\advance \IIInum by 1
            {\the\Inum.\the\IInum.\the\IIInum. }}
\def\IV{\global\advance \IVnum by 1
            {\the\IVnum. }}
\slugcomment{Submitted to ApJ January 2007}%; accepted February 2007}

\shorttitle{Thermal and Nonthermal Radiation in SNRs}
\shortauthors{Ellison et al.}

\begin{document}

\title{Particle Acceleration in Supernova Remnants and the Production of
Thermal and Nonthermal Radiation}

\author{Donald C. Ellison} 
\affil{Physics Department, North Carolina State University, Box 8202, Raleigh, NC
27695, U.S.A.,  {\tt don\_ellison@ncsu.edu}}
\author{Daniel J. Patnaude} 
\affil{Harvard Smithsonian Center for Astrophysics, 60 Garden St.,
Cambridge, MA 02138, U.S.A., {\tt patnaude@head.cfa.harvard.edu}}

\author{Patrick Slane} 
\affil{Harvard Smithsonian Center for Astrophysics, 60 Garden St.,
Cambridge, MA 02138, U.S.A., {\tt slane@cfa.harvard.edu}} 

\author{Pasquale Blasi}
\affil{INAF/Istituto Nazionale di
  Astrofisica, Osservatorio Astrofisico di Arcetri, Largo E. Fermi 5,
  I-50125, Firenze, Italy, {\tt blasi@arcetri.astro.it}}

\author{Stefano Gabici}
\affil{Max-Planck-Institut fuer Kernphysik, Saupfercheckweg 1, 69117
  Heidelberg, Germany {\tt Stefano.Gabici@mpi-hd.mpg.de}}

\begin{abstract}
Energetic particle production in supernova remnants (SNRs) through
diffusive shock acceleration has long been suggested as a mechanism
by which the bulk of cosmic rays with energies less than $\sim
10^{15}$~eV are formed.  If highly efficient, this process can have
a significant effect on the X-ray emission from SNRs as well as their
dynamical evolution.  Here we investigate the expected modification
to the thermal X-ray emission
from the forward shock region of young SNRs in which efficient particle
acceleration is occurring. Using hydrodynamical simulations for a range
of ambient density and magnetic field values, we produce spectra for
both the thermal and nonthermal emission components of the postshock
gas. For a given ambient density and explosion energy, we find that the
position of the forward shock at a given age is a strong function of the
acceleration efficiency, providing 
a strong signature of cosmic ray production.
Using an approximate treatment for the ionization state of the plasma,
appropriate for the range of models considered, we investigate the
effects of slow vs. rapid heating of the postshock electrons on the
ratio of thermal to nonthermal X-ray emission at the forward shock. 
We also investigate the effects of magnetic field strength on the
observed spectrum for efficient cosmic ray acceleration. The primary
effect of a large field, aside from an overall increased flux for
higher fields, is a considerable flattening of the nonthermal spectrum
in the soft X-ray band. Spectral index measurements from X-ray
observations may thus be used as indicators of the postshock magnetic
field strength.  The predicted gamma-ray flux from \IC\ (IC)
scattering and neutral pion ($\pi^0$) decay is strongly affected by the
ambient conditions.  For the particular parameters used in our examples,
the IC emission at $E \sim 1$~TeV exceeds that from $\pi^0$ decay,
although at lower energies of several GeV and at higher energies $>$TeV,
this trend is reversed for cases of high ambient density.  More
importantly, for high magnetic fields, 
we find that radiation
losses, combined with evolutionary effects, produce a steepening of the
electron spectrum over a wide energy range and results in a severe
flattening of the IC spectral shape, making it more difficult to
differentiate from that due to pion decay.
\end{abstract}


\section{INTRODUCTION}
The production of \rel\ electrons is beyond doubt in young supernova
remnants (SNRs) and it is believed that remnants simultaneously produce
\rel\ ions, i.e., cosmic rays (CRs), although the evidence for this is less
direct. 
SNRs also produce thermal X-ray line emission, and this contains a vast
amount of information about the composition and ionization state of the
gas that is absent in continuum emission.  The premise of this paper is
that shocks in SNRs accelerate ions {\it efficiently} via diffusive
shock acceleration (DSA) and that this particle acceleration influences
the evolution of the remnant, and in particular, the {\it thermal} X-ray
emission, in important and predicable ways. In the current era of high
resolution, high energy astronomy, it has become critical that \SC,
broad-band models of SNR radiation under the assumption of efficient
particle acceleration be developed.
Towards this goal, we present a \SC\ model of thermal and \NT\ X-ray
emission in SNRs. The model includes the acceleration and feedback of
CRs on the SNR evolution and on the structure and thermal properties of
the shocked gas in the interaction region between the forward and
reverse shocks.  In this paper, the first in a series, we investigate
the dynamical and spectral evolution of the forward shock region only,
where the vast majority of the cosmic ray production is expected to
occur. In subsequent papers we will treat results for the ejecta
component and provide a more complete treatment of the ionization within
the SNR.

Radiation from shell-like SNRs consists of thermal emission from
shock-heated gas and non-thermal emission from shock accelerated
particles.\footnote{We do not consider emission from heated dust or from
a compact object in this paper.}
X-rays consist of a thermal component which contains emission from both
shock-heated ejecta and circumstellar medium (CSM).
Since the ejecta and CSM generally have very different
compositions, the separation of these two components observationally is
often possible.
In this paper we concentrate on thermal emission from the shocked
circumstellar medium (CSM) material in order to highlight the physical
effects we wish to emphasize without the additional complications
introduced by the complex composition and ionization state of the
shocked ejecta.

A number of SNRs also show non-thermal X-ray emission 
and some (e.g., RX J1713.7-3946) even show non-thermal
X-rays with no discernable thermal component \citep[][]{Slane99}.
Non-thermal X-rays are believed to be \synch\ radiation from shock
accelerated TeV electrons.
Radio emission has long been observed from SNRs in the form of \synch\
radiation from \rel\ electrons
accelerated at the outer blast wave.
In addition, higher energy photons at GeV--TeV energies (either \IC\
radiation from electrons or \pion\ emission from ions) have been
detected from some remnants, particularly by HESS and CANGAROO
\citep[e.g.,][]{AharonianNature2004,TanimoriEtal1998} and now MAGIC
\citep[e.g.,][]{MAGIC_J1813_2006}, and soon VERITAS
\citep[][]{LeBohec2006}.

If the particle acceleration process is efficient DSA, as is now
becoming more generally accepted, the \NL\ theory makes clear
predictions for the underlying electron and ion spectra which produce
the broad-band photon emission \citep[e.g.,][]{JE91,BaringEtal99}. While
a number of authors have calculated the continuum emission from \synch,
\IC, and \pion\ emission \citep*[e.g.,][]{ESG2001,BKV2002}, there have
been, to our knowledge, no \SC\ calculations of the thermal and \NT\
X-ray emission in SNRs.
The most complete attempt to do this that we are aware of is that of
\citet*{DEB2000}, who used a self-similar analytic model of the SNR
hydrodynamics with input from a \NL\ calculation of DSA, which modified
the thermal properties of the interaction region. This model did not,
however, \SCly\ calculate the SNR evolution or include the \NT\ \synch\
component \SCly. Earlier work by \citet{DorfiBohr93} \citep[see
also][]{ch83,Heavens84,BoularesCox88}, clearly showed the importance of
including \NL\ particle acceleration in SNR models but did not attempt
\SC\ calculations of thermal and \NT\ emission in the X-ray band.

\newlistroman

In this paper, we have incorporated a non-equilibrium ionization (NEI)
model of X-ray line and continuum emission into a spherically symmetric
hydrodynamic simulation of SNRs.  The hydrodynamics include
efficient, \NL\ DSA and the effects of this particle acceleration on the
shocked thermal plasma, as well as on the structure and evolution of the
remnant.
Our model includes:
\listromanDE shock heating at both the
forward and reverse shocks;
\listromanDE \NL\ particle acceleration at the forward shock;
\listromanDE a consistent modeling of the evolution of the material in
the interaction region between the forward and reverse shocks; and
\listromanDE the calculation of full electron and proton distribution
functions which allow the \NT\ \synch\ contribution to the X-ray
emission to be determined \SCly\ with the thermal emission.  The thermal
emission is calculated for an optically thin plasma in nonequilibrium
ionization with negligible cooling.

Efficient DSA predicts greater compression ratios and lower plasma
temperatures than produced in shocks with little or no particle
acceleration.  These two effects will influence the X-ray emission from
the shocked CSM where DSA is expected to be efficient.  The changes in
SNR evolution produced by efficient DSA at the forward shock (FS) modify
the thermal emission from the shocked ejecta even if DSA is absent at
the reverse shock (RS). While these changes may be small, the \NT\ \syn\
emission from the FS may produce large changes in the observed ejecta
emission through line-of-sight effects even without DSA at the RS.

In $\S$ 2 we outline our model. We use the results from time-dependent 
hydrodynamical models with efficient DSA as inputs to a NEI calculation. We 
discuss our approach to the problem of nonequilibrium ionization in $\S$ 2
as well. We present our results in $\S$ 3 and discuss the quantitative
and qualitative effects of DSA on thermal X-ray emission in $\S$ 4. Finally,
in $\S$ 4 we present and discuss the limitations of our approach and 
present possible solutions to these limitations as well as future directions.

\newlistroman

\section{CR-HYDRO MODEL WITH X-RAY EMISSION}
Our model consists of three main parts:
\listromanDE the hydrodynamics of the SNR coupled to \NL\ DSA (our
so-called CR-hydro model);
\listromanDE a calculation of X-ray \synch\ emission from shock
 accelerated electrons; and
\listromanDE a \NEI\ (NEI) calculation of thermal X-ray emission using the
output of the CR-hydro model. %; and

\begin{figure}        % Fig 1
\epsscale{0.55} \plotone{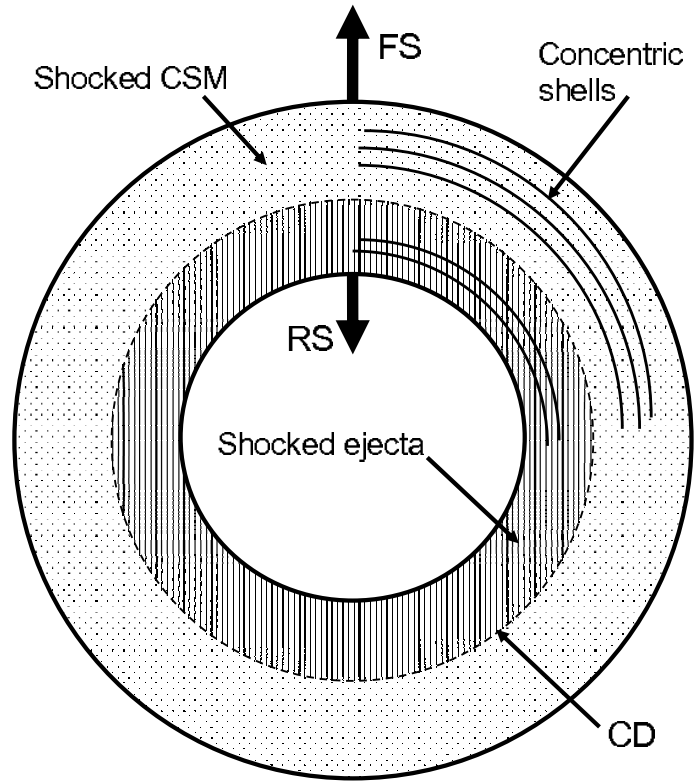} %%SNR_sketch_3.eps}
\caption{Spherical SNR model showing the forward shock (FS), reverse
  shock (RS), and contact discontinuity (CD). The CD separates the
  shocked circumstellar medium (CSM) from the shocked ejecta. The RS is
  an inward directed shock even though it moves outward during the early
  stages of the SNR evolution. As the forward and reverse shocks
  overtake new material, the cosmic-ray population is determined in
  concentric shells (indicated schematically) that evolve as the SNR ages.
\label{SNR_sketch}}
\end{figure}

% ttt111
\begin{table}
\begin{center}
\caption{Parameters for models shown in
Figures~\ref{vary_inj}--\ref{fp_TP_NL} and
Figures~\ref{line_Coul}--\ref{IC_pion}. In all models $\Mej=1.4\,\Msun$,
$\tSNR=500$\,yr, $\dSNR=1$\,kpc, and an exponential ejecta density
profile is used. For all models except D2, $\EnSN=10^{51}$\,erg. For D2,
$\EnSN=1.4\xx{51}$\,erg.}  
\vskip6pt
\begin{tabular}{crrrrrrrrrrr}
\tableline
\tableline
Model\tablenotemark{a}
\tablenotetext{a}{Models A, B, and C are shown in Figures~\ref{line_eff},
  \ref{line_density}, and \ref{line_B}, respectively. Models D1 and D2
  are shown in Figure~\ref{vary_inj}} 
&$\denH$ 
&$B_0$
&$\Inj$ 
&$\EffRel$ 
&$\EffEsc$\tablenotemark{b}
\tablenotetext{b}{This is the percent of energy flux which escapes
  upstream of the FS at $\tSNR$.} 
&$\Rtot$\tablenotemark{c}
\tablenotetext{c}{This is the FS compression ratio at $\tSNR$.}
&$B_2$\tablenotemark{d} \tablenotetext{d}{This is the magnetic field
immediately behind the FS at $\tSNR$.}  
&$\RFS$\tablenotemark{e} \tablenotetext{e}{Radius of the FS at $\tSNR$.}
&(\synTherm)\tablenotemark{f}\tablenotetext{f}{These values are summed over the
  $0.4-10$\, keV energy range and are for the \CH\ case.}
&\pionICfr
\\
&[\pcc]
&[\muG]
&
&[\%]
&[\%]
&
&[\muG]
&[pc]
\\
\tableline
A1
&0.1
&15
&6
&0
&0
&4
&60
&5.0
&0
&\dots
\\
A2
&0.1
&15
&3.91
&10
&1.6
&4.3
&65
&4.9
&4.68
&0.048
\\
A3
&0.1
&15
&3.62
&50
&16
&6.5
&98
&4.7
&13.8
&0.20
\\
\tableline
B1
&0.01
&15
&3.58
&36
&7.5
&5.5
&83
&6.6
&3900
&0.010
\\
B2\tablenotemark{g} \tablenotetext{g}{Models B2 and C2 are identical.}
&0.1
&15
&3.73
&36
&9.4
&5.5
&83
&4.8
&13.5
&0.077
\\
B3
&1
&15
&3.82
&36
&11
&5.6
&84
&3.4
&0.44
&0.65
\\
\tableline
C1
&0.1
&3
&3.752
&36
&11
&5.6
&17
&4.9
&1.5
&0.027
\\
C2\tablenotemark{g}
&0.1
&15
&3.73
&36
&9.4
&5.5
&83
&4.8
&13.5
&0.077
\\
C3
&0.1
&60
&3.2
&36
&19
&5.2
&310
&4.7
&22.7
&1.1
\\
\tableline
D1
&0.1
&15
&3.3
&63
&23
&7.7
&116
&4.6
&\dots
&\dots
\\
D2\tablenotemark{h} \tablenotetext{h}{This model has
  $\EnSN=1.4\xx{51}$\,erg and should be compared to model A1 with
  $\EnSN=1\xx{51}$\,erg.}  &0.1 &15 &3.3 &64 &24 &7.8 &120 &5.0 &\dots
  &\dots 
\\
\tableline
\end{tabular}
\end{center}
\label{table1}
\end{table}
%above ttt111

\subsection{CR Hydrodynamics and Electron Spectra}
We calculate the hydrodynamic evolution of a SNR with a radially
symmetric model described in detail in \citet{EC2005} and references
therein. For reference, we reproduce the geometry of the simulation in
Figure~\ref{SNR_sketch}.
The current model couples efficient DSA to the hydrodynamics and differs
from that described in \citet{EC2005} in that we have replaced the
algebraic model of DSA of \citet{BE99} with the more accurate \SA\ model
of \citet*{BGV2005}.\footnote{The work of \citet{BGV2005}, which
is based on the approximate solution of \citet{Blasi2002,Blasi2004},
showed that multiple solutions could be eliminated by
relating the injection momentum to the thermal momentum of the shocked
particles. Recently, an exact solution for arbitrary conditions has been
presented by \citet{AB2005,AB2006}.}
Given an injection parameter, $\Inj$ \citep[this is $\xi$ in
equation~(25) in][]{BGV2005}, the \SA\ model solves the \NL\ DSA problem
at each time-step of the hydro simulation using the shock speed, shock
radius, ambient density and temperature, and ambient magnetic field
determined in the simulation.
With the accelerated particle distribution, an
effective ratio of specific heats is calculated and used in the
hydrodynamic equations, completing the coupling between the two
\citep*[see][ for a fuller discussion]{EDB2004}.

At any given time, the CR-hydro calculation returns a set of
concentric shells of shocked material, both shocked ejecta and shocked
CSM, separated by a \CD\ (CD). The hydrodynamics are calculated in
Lagrangian mass coordinates so the amount of material in a mass shell
remains constant as the volume of a shell is adjusted in response to the
changing pressure.  New shells of shocked gas are added as the forward
and reverse shocks overtake fresh material.

Particle acceleration influences the SNR {evolution} since \rel\
particles produce less pressure for a given energy density than do
\nonrel\ particles. In addition, particles at the maximum energy
``escape'' upstream from the shock system during acceleration and carry
away energy and pressure, thus further softening the equation of
state. This escape is included implicitly in the semi-analytic DSA
calculation \citep[see][]{BE99} and is distinct from the diffusion of
particles into and out of a particular shell.
Since we
use Lagrangian coordinates we ignore the diffusion of
particles into and out of a shell after the initial acceleration. 
This is
a reasonable approximation for low energy particles, but becomes less
accurate as the particle energy and diffusion lengths increase.

The softening of the 
equation of state means that compression ratios well in excess of four
can be produced in non-radiative, collisionless shocks further modifying
the SNR evolution \citep[e.g.,][]{Eichler84}.
The nature of \NL\ DSA, on the other hand, uniquely predicts that a
concave superthermal particle spectrum, hardening at higher momenta,
will be produced. Here the energy that goes into the highest momenta
particles comes from the shock-heated thermal population.  
Thus, the fundamental changes that occur with efficient DSA are larger
compression ratios, lower temperatures of the shocked gas
(compared to shocks without efficient DSA), and concave energetic
particle spectra.\footnote{The concave curvature in the superthermal
portions of the spectra in the middle panel of Figure~\ref{fp_TP_NL} is
present but small. The curvature becomes more pronounced with greater
acceleration efficiency \citep*[see Fig.~1 in][]{EBB2000} and may
actually have been detected in SNR spectra
\cite[e.g.,][]{RE92,JonesEtal2003,AllenEtal2005}.}

\begin{figure}        % Fig 2
\epsscale{.60} \plotone{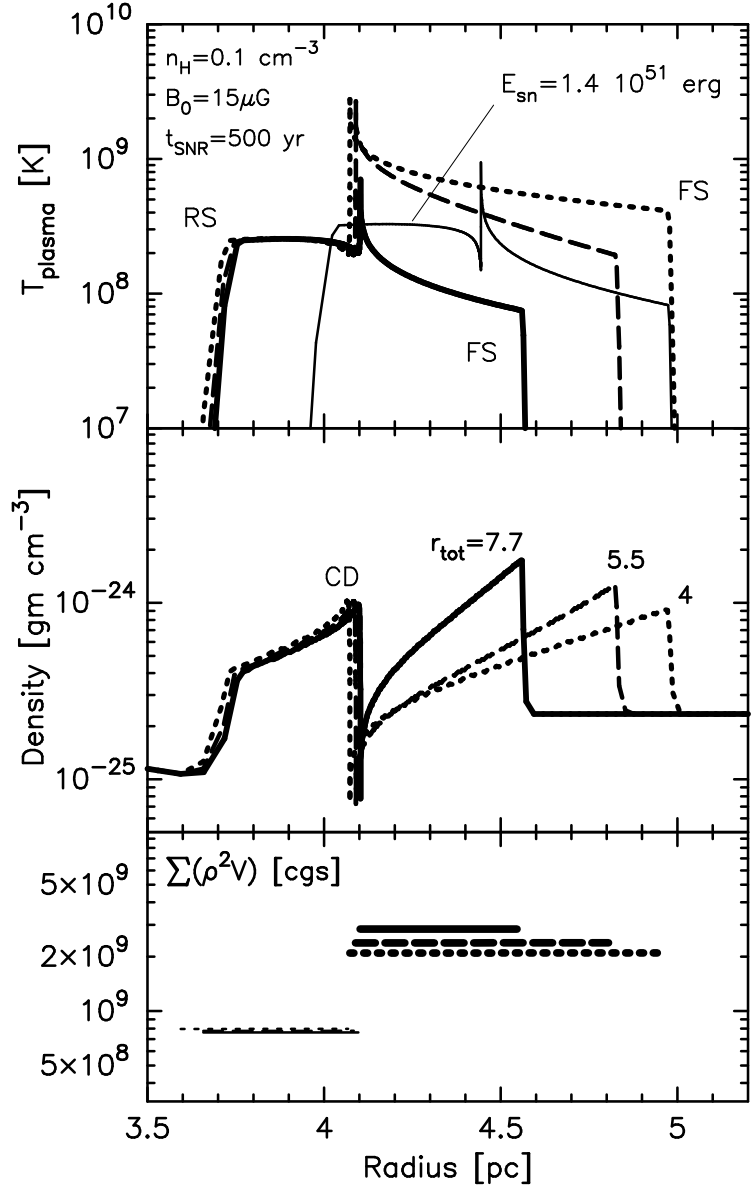} %%3stack_vary_inj_n0.1_age500_pc.eps}
\caption{The top and middle panels show radial distributions of the
plasma temperature and density for SNRs with different shock
acceleration efficiencies at the FS. The bottom panel shows
$\sum{\rho^2V}$ where each horizontal line indicates the value summed
over the shocked ejecta or shocked CSM regions. In all panels, the
heavy-weight solid curves (model D1 in table) are for efficient DSA
($\EffRel \simeq 63\%$), the dashed curves (model B2) are for moderate
DSA ($\EffRel \simeq 36\%$), and the dotted curves (model A1) are for
\TP\ ($\EffRel \simeq 0$). In the top panel, the light-weight solid
curve (model D2) is an efficient acceleration case ($\EffRel \simeq
63\%$) where $\EnSN$ has been chosen to produce the same FS radius as
the TP example with $\EnSN=10^{51}$\,erg In all cases, no DSA occurs at
the RS.
\label{vary_inj}}
\end{figure}

An important question for SNRs is whether or not particle
acceleration occurs at the RS. While it is likely that, due to
expansion, the magnetic field interior to the RS is too small to allow
acceleration, there have been reports of radio
\citep[][]{GotthelfEtal2001,DeLaneyEtal2002} and even X-ray \syn\
emission \citep{RhoEtal2002} at the RS in some SNRs.  As discussed in
\citet*[][]{EDB2005}, confirmation of particle acceleration at the RS
will indicate strong magnetic field amplification and will have
far-reaching consequences.  For now, however, we assume there is no
appreciable acceleration at the RS, but note that the large changes in
the density and temperature that occur with efficient DSA may make it
possible to use observations of thermal X-rays to set limits on the
amount of acceleration.

We use as inputs to our thermal models a set of hydrodynamic models in
which we vary the ambient magnetic field $B_0$, the ambient density
$n_H$, and the acceleration efficiency $\EffRel$.  The acceleration
efficiency, $\EffRel$, is defined as the percentage of energy flux
crossing the shock (in the shock rest frame) that ends up in \rel\
particles, and it is uniquely determined by $\Inj$. The acceleration
efficiency includes particles in the particle distributions downstream
from the shock as well as the high-energy particles that escape upstream
from the shock.
All models are initialized in a similar way typical of Type Ia
supernovae: we use an exponential density profile
\citep[e.g.,][]{Dwarkadas2000} for the SNR ejecta, an ejecta mass $\Mej
= 1.4\,\Msun$, and an ejecta kinetic energy $\EnSN = 10^{51}$ erg.  We
assume that the ambient magnetic field and density are uniform and
evolve the models for a time $\tSNR$, typically taken to be
500\,yr.\footnote{A remnant age of $\tSNR=500$\,yr is chosen mainly for
computational convenience although it allows comparison with some young
historical SNRs.}

\begin{figure}        % Fig 3
\epsscale{.60} \plotone{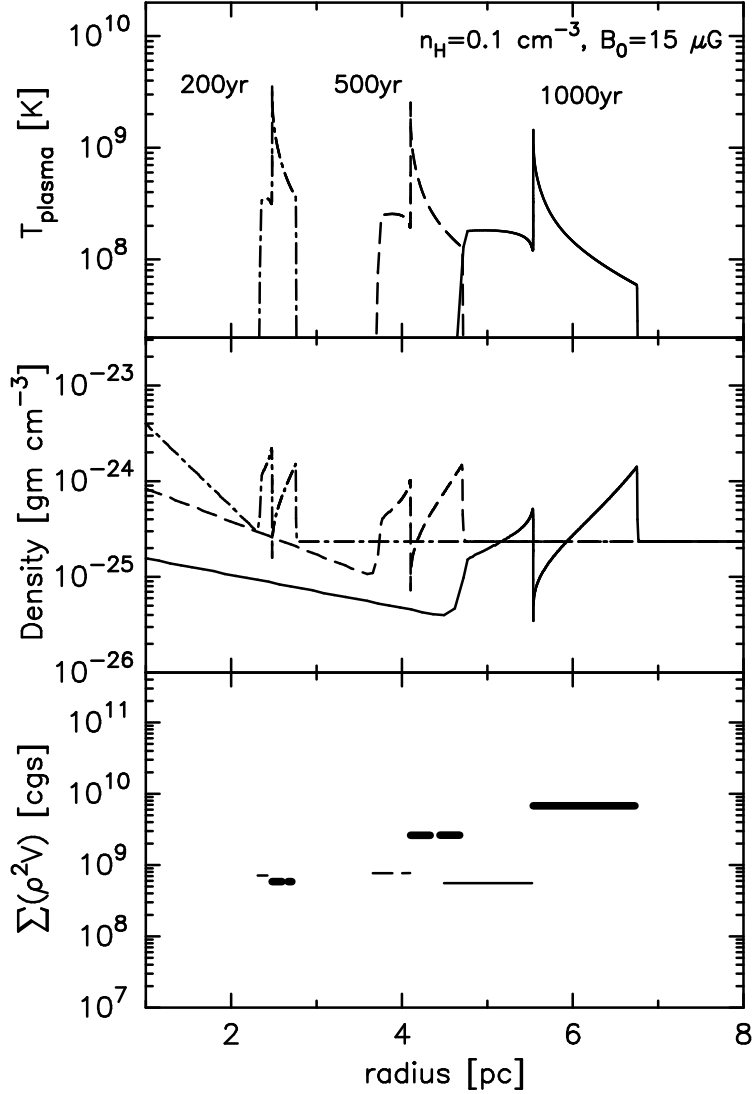} %%3stack_n0.1_50pc.eps} 
\caption{Temperature, density, and \SumR\ as in
  Figure~\ref{vary_inj}. The various sets of curves were calculated at
  $\tSNR=200$, 500, and 1000\,yr as indicated. In all examples,
  $\EffRel\sim 50\%$ at $\tSNR$. The dashed curves are model A3 in
  Table~1.
\label{den_const}}
\end{figure}

In Figure~\ref{vary_inj} we show the plasma temperature, density, and
volume emission measure (i.e., $\sum{\rho^2V}$) with varying injection
efficiencies for DSA.
In all cases, except for the light-weight curve in the top panel, we
take $\denH=0.1$\,\pcc, $B_0=15$\,\muG, $\tSNR=500$\, yr and
$\EnSN=10^{51}$\,erg. For the light-weight curve, we use
$\EnSN=1.4\xx{51}$\,erg with the same values for $\denH$, $B_0$, and
$\tSNR$. 
Other parameters are listed in Table~1 as models A1, B2, D1,
and D2.

The three sets of heavy-weight solid curves have strong DSA with
$\EffRel \sim 63\%$ at $\tSNR=500$\,yr.  The dashed curves are for a
moderate acceleration efficiency with $\EffRel \sim 36\%$ and the dotted
curves are the \TP\ limit with negligible energy flux in \rel\ particles
($\EffRel \sim 0$).
These results clearly demonstrate the two most important effects
efficient CR production has on the remnant dynamics: the plasma
temperature behind the FS (top panel) drops by nearly a factor of 10
between the efficient acceleration and \TP\ examples, and the shock
compression (middle panel) immediately behind the FS increases by a
factor $7.7/4 \sim 2$.\footnote{The compression ratios indicated in
Table~1 are low compared to what is possible in nonlinear DSA
\citep[see, for example,][ for extreme cases with $\Rtot \sim
100$]{EDB2005}. The modest values for $\Rtot$ obtained here result from
a combination of restricting the acceleration efficiency and assuming
large upstream fields. As described in \citet{EDB2005}, in the
approximations used in our CR-hydro model, large upstream fields result
in heating of the shock precursor and reduce the overall acceleration
efficiency and $\Rtot$.} %\DCE

The quantity $\sum{\rho^2 V}$ in the bottom panel (where $\rho$ is the
shocked plasma density and $V$ is the volume of the shocked plasma), is
similar to the emission measure (EM); the thermal X-ray emission is
proportional to $\sum{\rho^2V}$.  The light-weight bar is
$\sum{\rho^2V}$ for the region between the reverse shock and the \CD,
and the heavy-weight bar is for the region between the forward shock and
the \CD.

An important consequence of efficient DSA is that it results in a
smaller forward shock radius at a given age as compared to the \TP\
limit. 
A comparison of the FS positions of the
heavy-weight solid and dotted curves in Figure~\ref{vary_inj} shows that
the FS radius drops by a factor of $4.6/5 \sim 0.9$ between the \TP\ and
highly efficient acceleration cases with a corresponding drop in the
FS/CD ratio.
Such an effect is observed in Tycho's SNR \citep{WarrenEtal2005},
providing
strong evidence for efficient particle acceleration in this young remnant.
The light-weight solid curve in the top panel has strong DSA ($\EffRel
\sim 63\%$), but the
supernova explosion energy has been increased from $10^{51}$\,erg
(heavy-weight curves) to $1.4\xx{51}$\, erg to obtain the same FS
radius as the \TP\ case. If efficient DSA occurs and is ignored,
systematic errors in estimates for $\EnSN$ will occur.

In Figure~\ref{den_const} we show the shocked plasma temperature,
density, and $\sum{\rho^2 V}$ at $\tSNR=200$, 500, and 1000\,yr. These
are efficient forward shock acceleration cases with $\EffRel \sim 50\%$
at $\tSNR$.
At $\tSNR=200$\,yr, $\sum{\rho^2V}$ is slightly higher at the RS, but by
1000\,yr, $\sum{\rho^2V}$ is almost an order of magnitude higher in the
FS. This occurs, of course, because the FS continues to sweep up CSM
material as the remnant expands.

\begin{figure}        % Fig 4
\epsscale{.60} \plotone{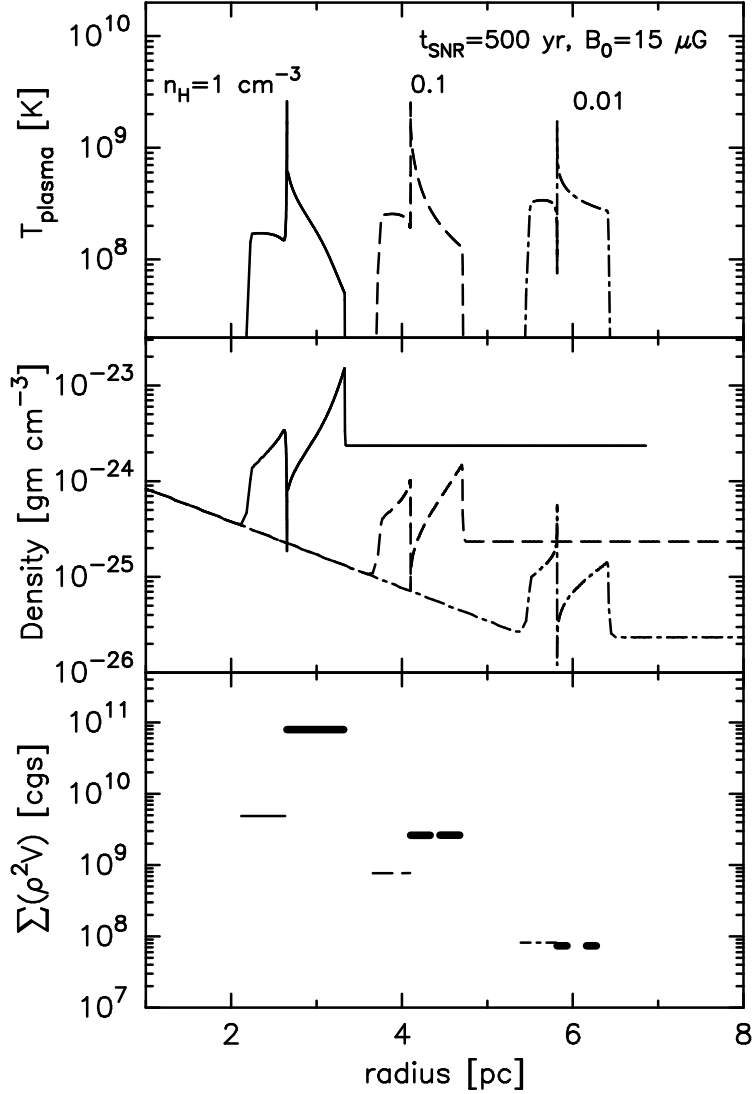} %%3stack_age500_50pc.eps} 
\caption{Temperature, density, and \SumR\ as in
  Figure~\ref{vary_inj}. The various sets of curves were calculated with
  different $\denH$ as indicated. In all cases, $\EffRel\simeq
  50\%$. The dashed curves are model A3 in Table~1.
\label{age_const}}
\end{figure}

\begin{figure}        % Fig 5
\epsscale{.60} \plotone{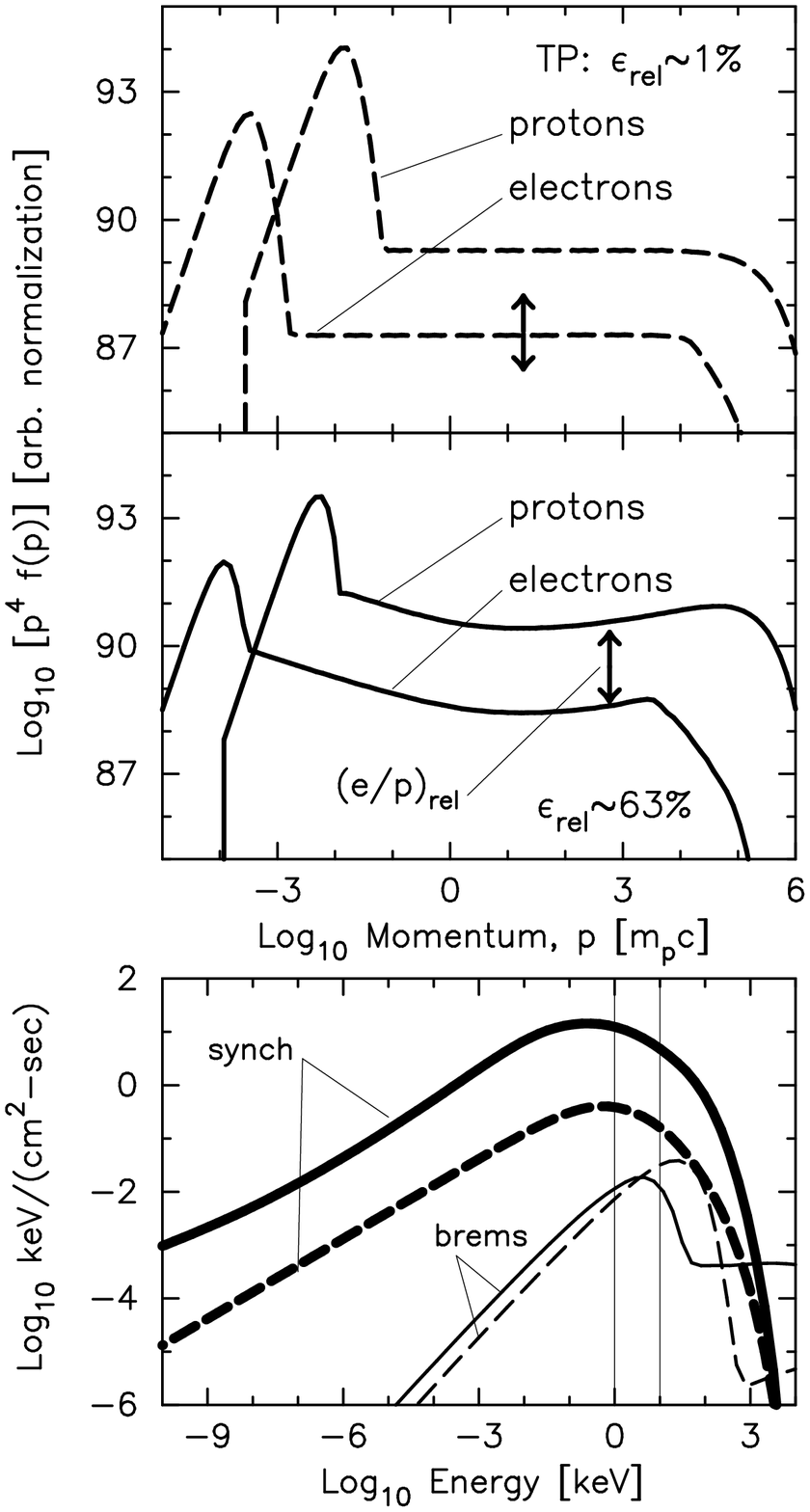} %%fp_phot_TP_NL_pc.eps} 
\caption{The top two panels show electron and proton spectra, $p^4
f(p)$, for TP DSA ($\EffRel \sim 1\%$, dashed curves, $\sim$ model A1)
and efficient DSA ($\EffRel \sim 63\%$, solid curves, model D1). The
bottom panel shows \syn\ and \brem\ spectra for these two cases: solid
curves - efficient DSA, dashed curves - TP.  In all cases,
$B_0=15$\,\muG\ and $\denH=0.1$\,\pcc.  There is a large variation in
the synch/brem ratio in the X-ray range between these two cases. The
up-down arrow in the top panel indicates that the normalization of the
power law, relative to the thermal distribution, is arbitrary in the
\TP\ case other than that the power law is low enough to contain an
insignificant fraction of the total energy.
The up-down arrow in the middle panel shows the arbitrary parameter
$\epRel=0.01$, typical of the ratio observed in galactic cosmic
rays. The thermal electron distributions shown assume instant
temperature equilibration with protons behind
the shock.
\label{fp_TP_NL}}
\end{figure}

Figure~\ref{age_const} shows a similar set of curves, also for $\EffRel
\sim 50\%$, only now the age
is fixed at 500\,yr and $\denH$ is varied as shown.  As expected,
$\sum{\rho^2V}$ at the forward shock scales roughly as $\denH^2$. At the
RS, $\sum{\rho^2V}$ also shows a correlation with $\denH$ but it is
considerably weaker than $\denH^2$.

In addition to modifying the {evolution} and the temperature of the
shocked gas, changes in the compression of the fluid should result in
changes in the compression of the magnetic field, implying that the
morphology and intensity of \syn\ emission from \rel\ electrons will
vary strongly with the efficiency of DSA and with the orientation and
strength of the magnetic field.\footnote{We note that the orientation of
the magnetic field is not included explicitly in our model. As discussed
in \citet{EC2005}, for the purposes of calculating the evolution of the
field, we assume the shocked field is fully turbulent.}
The magnetic field in the CR-hydro model is exactly as described in
\citet{EC2005} and we emphasize that the DSA model used here does not
explicitly include the magnetic field.
What this means is that instead of including the magnetic field \SCly\
in determining the shock structure, the far upstream field strength,
$B_0$, is used to calculate a rate of \alf\ heating in the shock
precursor \citep[as described in][]{BE99} which in turn modifies the
subshock compression and overall acceleration efficiency.  An ad hoc
field compression model is assumed to determine the downstream field
\citep[e.g.,][]{EC2005} and this compressed field is used to calculate
\syn\ emission.

Significantly, our model does not include magnetic field amplification
even though this is now believed to be an important effect in SNR shocks
\citep[e.g.,][]{VL2003,VBK2005a}. Field amplification, as opposed to
simple compression, comes about when backstreaming energetic particles
transfer energy to magnetic turbulence to create a \NL\ situation where
$\Delta B/B \gg 1$ \citep[][]{BL2001}.  Efforts are now underway to
incorporate this process into \NL\ models of DSA
\citep*[i.e.,][]{AB2006,VEB2006,BAC2006}. We do not attempt to model
such amplification here, but \citep[in a fashion similar to, for
example,][]{VBK2005a}, we investigate the effects of the associated
strong postshock magnetic fields by considering the case with a strong
(60 $\mu$G) upstream field which, upon strong compression under
conditions of efficient particle acceleration, in some ways mimics the
amplification process.

\newlistroman

In Figure~\ref{fp_TP_NL} we show electron and proton spectra for two
cases, one with \TP\ acceleration, ($\EffRel\sim 0$; dashed
curves, top panel) and one with efficient DSA, ($\EffRel \sim 63\%$;
solid curves, middle panel).
For this example, the specific parameters are not important other than
to emphasize that the only difference in input between the two models is
the injection efficiency at the forward shock $\Inj$.  The spectra shown
are integrated over the region between the CD and FS.  Both the spectral
curvature and the shift of the thermal peak to lower momenta with
efficient DSA are evident.
In the bottom panel of Figure~\ref{fp_TP_NL} are the \synch\ and \brem\
spectra
generated by the electron distributions. The two vertical lines show
the 1-10 keV range relevant for X-ray observations. 
The important points to be made here are:

\listromanDE The spectral curvature in the \synch\ emission is far too
small to show up in the limited X-ray range, although it is important
for matching radio to X-ray fluxes since \NL\ effects increase the
X-ray/radio ratio significantly
\citep[][]{BaringEtal99,ESG2001,BKV2002};

\listromanDE In the models shown, the \synch\ spectral shape in the
critical 1-10 keV range is determined by the shape of the turnover in
the electron distribution near the maximum momentum. 
The momentum and shape of this turnover is determined by
$\Acut$,\footnote{The parameter $\Acut$ is described in detail in
\citet{EDB2004} (see eq.~2 in that paper). Simply stated, when
$\Acut=1$, an exponential turnover is imposed on the particle spectra
near $\pmax$. This turnover can be important for modeling \syn\ X-rays
but the actual shape of the turnover can't be determined with the
semi-analytic code used here.}
radiation losses (\syn\ and \IC\ losses off the
primordial background radiation) and adiabatic expansion as well as the
maximum momentum, $\pmax$, the shock can produce.
Both the rate of radiation losses and $\pmax$ 
depend on the momentum dependence of the diffusion coefficient,
$\kappa$, near $\pmax$, which is unknown. To our knowledge, all serious
models of SNRs currently in use, including ours, assume that $\kappa
\propto p$ at $\pmax$, but this need not be the case
\citep*[see][]{AB2006,VEB2006,BAC2006}.
If $\kappa$ near $\pmax$ has a momentum dependence other than $\propto
p$, the shape of the \synch\ emission near 1 keV can be substantially
modified. 

\listromanDE The \brem\ continua without emission lines in the bottom
panel (light-weight curves) illustrate the effects on the thermal
emission.  First, note that the particle spectra in
Figure~\ref{fp_TP_NL} are ``complete'' in that they contain all of the
mass in the shocked gas. For this example, we assume the ambient plasma
is fully ionized hydrogen and helium with 10\% helium by number and that
$T_{2e} = T_{2p} = T_{2\alpha}$, that is, the shocked plasma of
electrons, protons, and $\alpha$-particles comes into equilibrium
immediately behind the shock.
In this example, the thermal continuum is dominated by the \synch\
[(heavy curve)/(light curve)] for both the TP and efficient
cases. However, the synch/brems ratio at 1 keV drops from $\sim 10^3$ in
the efficient case [(heavy solid curve)/(light solid curve)] to $\sim
50$ in the TP case [(heavy dashed curve)/(light dashed curve)].  The
shift to a lower electron temperature with efficient DSA is also evident
in the \brem\ curves.

Actual SNRs often show a mixture of thermal and nonthermal X-ray
emission \citep[Kepler's and Tycho's SNRs show a mixture, whereas in
G347.3$-$0.5, the X-ray emission is wholly nonthermal, e.g.,][]{Slane99}.  We
now describe how we use the above hydrodynamical models to 
calculate the thermal X-ray emission which is expected from cosmic-ray
modified shocks.

\subsection{NEI Model of Thermal Emission}

To properly predict the thermal X-ray emission produced by a
shock-heated gas, it is necessary to determine the ionization state for
each element in the gas. Particularly for the low densities found in
SNRs, this ionization state can differ significantly from that expected
for a gas in collisional ionization equilibrium at the associated
electron temperature of the postshock gas, $T_e$
\citep{HSC83,hamilton84}.  The approach to an equilibrium state for the
ionization depends upon both the electron density $n_e$ and the time
over which the gas has been ionizing (i.e. the time since it was
shocked). The non-equilibrium ionization (NEI) state for such an evolving
plasma has been calculated by a number of authors
\citep[e.g.,][]{hughes85,itoh79,borkowski01} the latter of whom have
implemented the calculation in a routine in the XSPEC software
package. Here the ionization state is calculated for given values of the
plasma temperature and so-called ionization timescale $\tau = n_e t$,
and the ionization values are then coupled to a plasma emissivity code
to produce the expected spectrum.

\begin{figure}        % Fig 6
\epsscale{0.85} \plotone{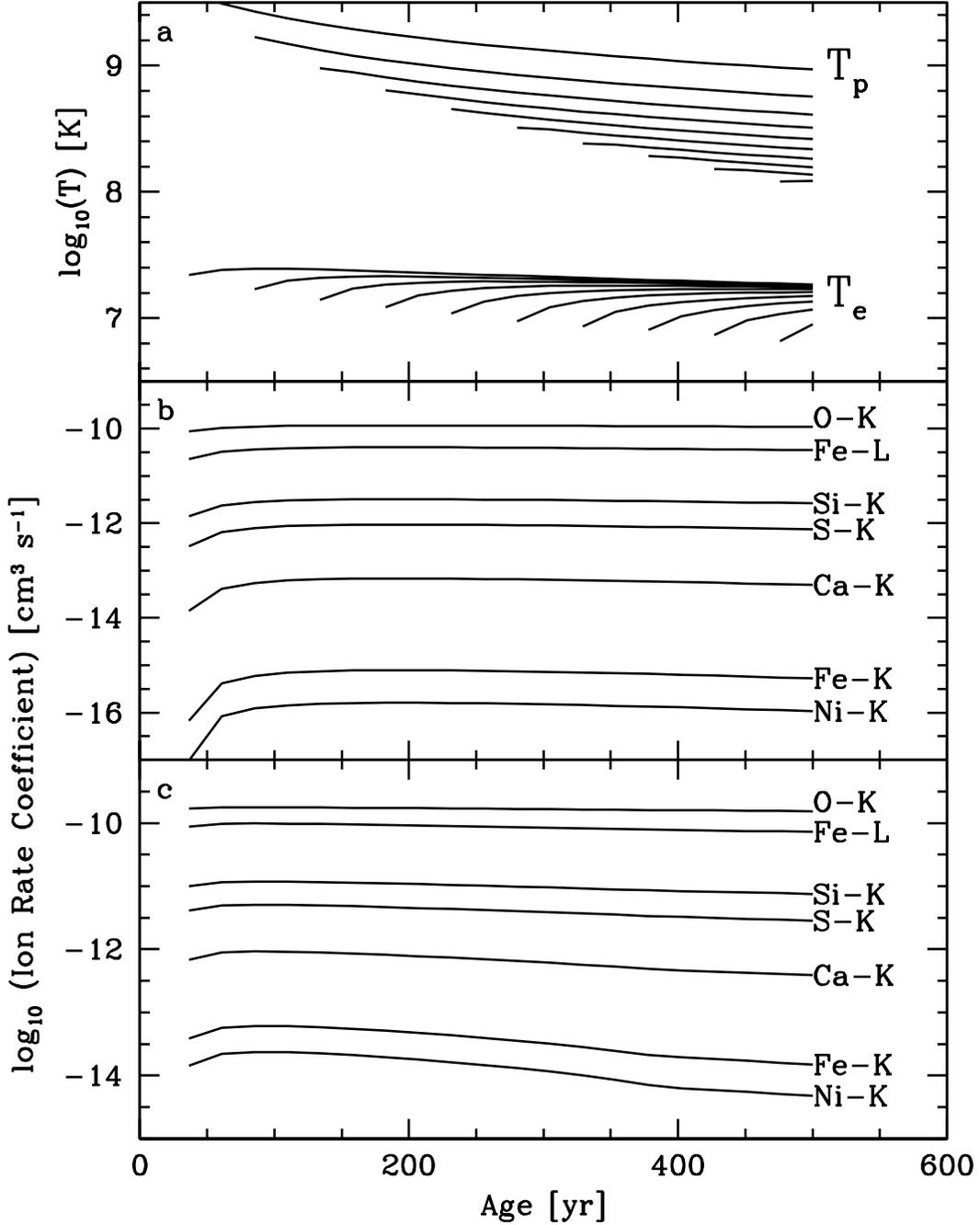} %%rates_Te_stack3.eps}
\caption{The top panel shows the electron and proton temperature
  evolution for various shells in the region between the CD and FS
  assuming heating by Coulomb collisions. The left end-point of each
  line indicates the SNR age at which the shell was formed. The lower
  two panels show the total K- and L-shell ionization rates for several
  X-ray emitting ions. Ions such as Mg-K and Ar-K are not shown but have
  similar profiles. In panels a and c, $\denH=1$\,\pcc, $B_0=15$\,\muG, 
  and $\EffRel \simeq 36\%$ (model B3, Table~1). In panel b,
  $\denH=0.1$\,\pcc, $B_0=15$\,\muG,  and $\EffRel \simeq 36\%$ (model
  B2).}
\label{tempevolve}
\end{figure}

To investigate the X-ray spectra from the shock-heated gas in an SNR
under conditions of efficient particle acceleration, we have coupled the
values for the evolution of the temperature and density in the postshock
gas, determined from hydrodynamical simulations, with the NEI model in
XSPEC. Specifically, for a remnant of a given age, we use the values of
$\tau$ and $T_e$, along with the volume emission measure $n_e n_H dV/(4
\pi D^2)$ for each radial shell from the simulation as input to the NEI
model ($D$ is the distance to the remnant). We consider two cases for
the heating of the electrons in the postshock region: \IE\ with the ion
temperature, and equilibration through Coulomb collisions
\citep{spitzer68}.  For both cases we include the effects of adiabatic
expansion on the temperature components.
We note that our line emissivity calculations assume a Maxwellian
distribution for the exciting electrons and ignore the possible
excitation by superthermal electrons or protons. When particle
acceleration is efficient, the low energy particles may, in fact, have a
distribution that differs somewhat from a Maxwellian although there is
no reason this difference should be significantly more than in the case
with inefficient acceleration.
More importantly, superthermal particles may contribute to the
ionization, particularly for the iron K-$\alpha$ line. These effects are
just starting to be considered \citep[e.g.,][]{Vink97,Porquet2001} and
are beyond the scope of this paper.

The ionization rate for a given species is highly dependent upon the
electron temperature.  As shown in the top panel of
Figure~\ref{tempevolve}, both the electron and ion temperatures evolve
as the SNR ages, due to Coulomb heating of the electrons and adiabatic
expansion (cooling) of the mass shell.  Therefore, the ionization state
at a particular time for any given shell is dependent upon the
ionization history in the shell, and is not a simple function of the
current temperature and summed $\tau$. While we defer a complete
treatment of the ionization directly in the hydrodynamical simulations
to a future paper, we find that there is a suitable parameter space over
which the ionization rates and temperatures are roughly constant in time
(except for when the shell is initially shock heated) and, thus, for
which the ionization rates for X-ray emitting ions are roughly constant
(panel b in Figure~\ref{tempevolve}). 
In such a case, the true ionization age of a mass shell is very close to
$n_e t_\mathrm{age}$, since the ionization versus time is roughly
constant. In the cases where the ionization rates are not nearly
constant with time 
(c.f., panel c in Figure~\ref{tempevolve}), 
the ionization rates for these states (i.e., Ni-K, Fe-K) are well below
the ionization rates for the dominant line emitters (e.g., Fe-L).
We note that in the case of shocked ejecta, this approximation will no
longer be valid, since there will be more free electrons per ion due to
the metal rich nature of the plasma, and this number will evolve as the
ejecta ionizes. We defer the treatment of the X-ray emission from the
reverse shock to a future publication.

XSPEC requires the following inputs: the electron temperature $T_e$, the
ionization age $n_e t$, the ion density $n_i$, and the shell volume. We
assume an arbitrary distance $D$ of 1 kiloparsec. The ion
density is related to the hydrodynamical gas density $\rho$ by $n_i$ =
$\rho/(\mu m_\mathrm{amu})$, where $\mu$ is the mean molecular weight
for a cosmic abundance plasma, taken here to be 0.6.  We then generate a
NEI model for each mass shell. The models are summed together to produce
an integrated thermal spectrum, such as that shown in
Figure~\ref{line_Coul}. These models can then either be folded through
an instrument response to produce a simulated spectrum, or added to the
nonthermal X-ray spectrum to produce a complete (thermal and nonthermal)
X-ray spectrum.\footnote{In the future, we intend to produce XSPEC table 
models from these combined spectra which can then be used to fit X-ray
observations of SNR.}

\section{RESULTS}
Calculations of SNRs including particle acceleration involve a number of
parameters.
For all of our results here, we 
 assume that the unshocked CSM is uniform (e.g., no pre-SN wind is
present), and 
\newlistroman
define the following environmental parameters for our spherically
symmetric calculation:
\listromanDE
initial kinetic energy of the ejecta material, $\EnSN$,
\listromanDE
mass of the ejecta, $\Mej$,
\listromanDE
age of the SNR, $\tSNR$,
\listromanDE
proton number density of the unshocked CSM, $\denISM$,
\listromanDE
magnetic field strength of the unshocked CSM, $B_0$, and
\listromanDE
composition of the CSM.

\newlistroman

Parameters and assumptions for the CR-hydro calculation include:
\listromanDE
the spatial profile for the ejecta material,
\listromanDE
the injection efficiency for DSA, $\Inj$,
\listromanDE the assumption for the scattering mean free path, i.e.,
$\lambda = \etamfp \rg$,
\listromanDE
the cutoff parameter for particle spectra near $\pmax$, $\Acut$, and
\listromanDE
the assumption for compression of magnetic field behind the shock.
The mean free path and $\etamfp$ only enter the calculation in the
determination of $\pmax$. The maximum momentum is determined when the
upstream diffusion length is equal to some fraction of the shock radius
or when the acceleration time equals the shock age, whichever gives the
lowest $\pmax$. For all of the models here, the fraction of the shock
radius that truncates the acceleration is $0.05$.
 
\begin{figure}        % Fig 7
\epsscale{.60} \plotone{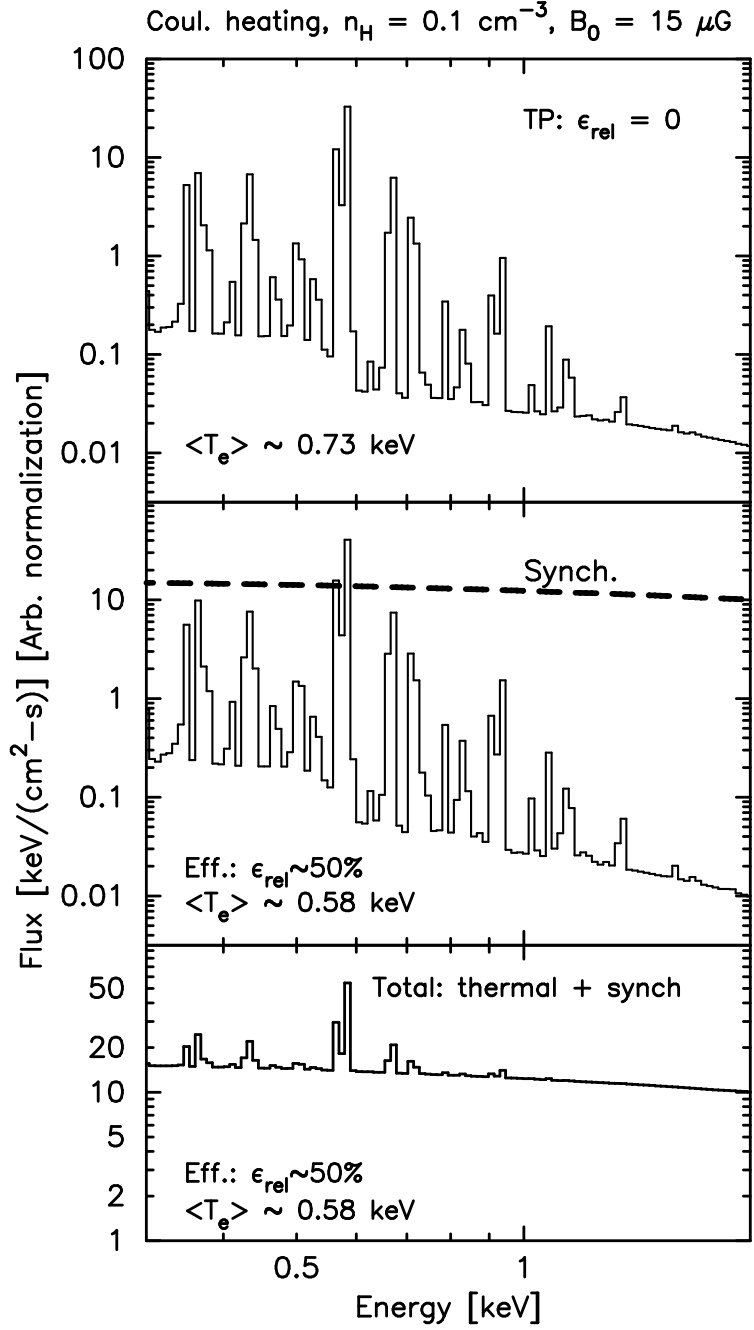} %%xray_line_TP_NL_Coul_50pc.eps} 
\caption{The top panel shows thermal X-ray emission with no DSA
  ($\EffRel=0$, model A1). The middle and bottom panels show spectra
  when the percentage of energy placed into \rel\ particles by the FS is
  $\sim 50\%$ (model A3 in Table~1). In all cases, only emission from
  the FS is shown and it is summed over the region between the FS and CD
  at $\tSNR=500$\,yr. The dashed curve in the middle panel is the \syn\
  flux from electrons accelerated by the FS and the bottom panel shows
  the total emission. For this example, \CH\ is assumed for electrons
  and the mean downstream temperatures are indicated.
\label{line_Coul}}
\end{figure}

\newlistroman

Parameters and assumptions for the NEI calculation include:
\listromanDE
the assumption for electron heating behind the shock,
\listromanDE the evolution of the plasma ionization, and
\listromanDE the chemical composition of the CSM.

In all of the examples that follow, we keep the following parameters and
assumptions fixed:
$\EnSN=10^{51}$\,erg,\footnote{We use $\EnSN=1.4\xx{51}$ in a single
  example shown in Figure~\ref{vary_inj}.}
$\Mej=1.4\,\Msun$,
an exponential density profile for the ejecta mass,
$\etamfp=1$,
$\Acut=1$,
and for the magnetic field configuration across the shock, we assume
that the field is fully turbulent upstream and, following
\citet{VBKR2002}, set the immediate downstream magnetic field
\begin{equation} \label{B_comp}
B_2= \sqrt{1/3 + 2 \Rtot^2/3}~B_0
\ ,
\end{equation}
where $\Rtot$ is the shock compression ratio.\footnote{The details of
  this assumption and for the evolution of $B$ behind the shock are
  given in \citet{EC2005}.}

An example product is given in Figure~\ref{line_Coul} where we show
emission in the X-ray energy range for a particular set of
parameters. For this illustration we have $\denISM=0.1$\,\pcc,
$B_0=15$\,\muG, and the results are shown at $\tSNR=500$\,yr. Here and
elsewhere, we only calculate line emission from the forward shock,
summed over the region between the FS and \CD, and do not adjust the
spectra for any type of instrument response.  We assume a solar
composition for the CSM and, most importantly, we assume that the
ionization age $\tau$ of the shocked CSM is given solely by the product
of the final electron density and the age of the mass shell.

\begin{figure}        % Fig 8
\epsscale{.85} \plotone{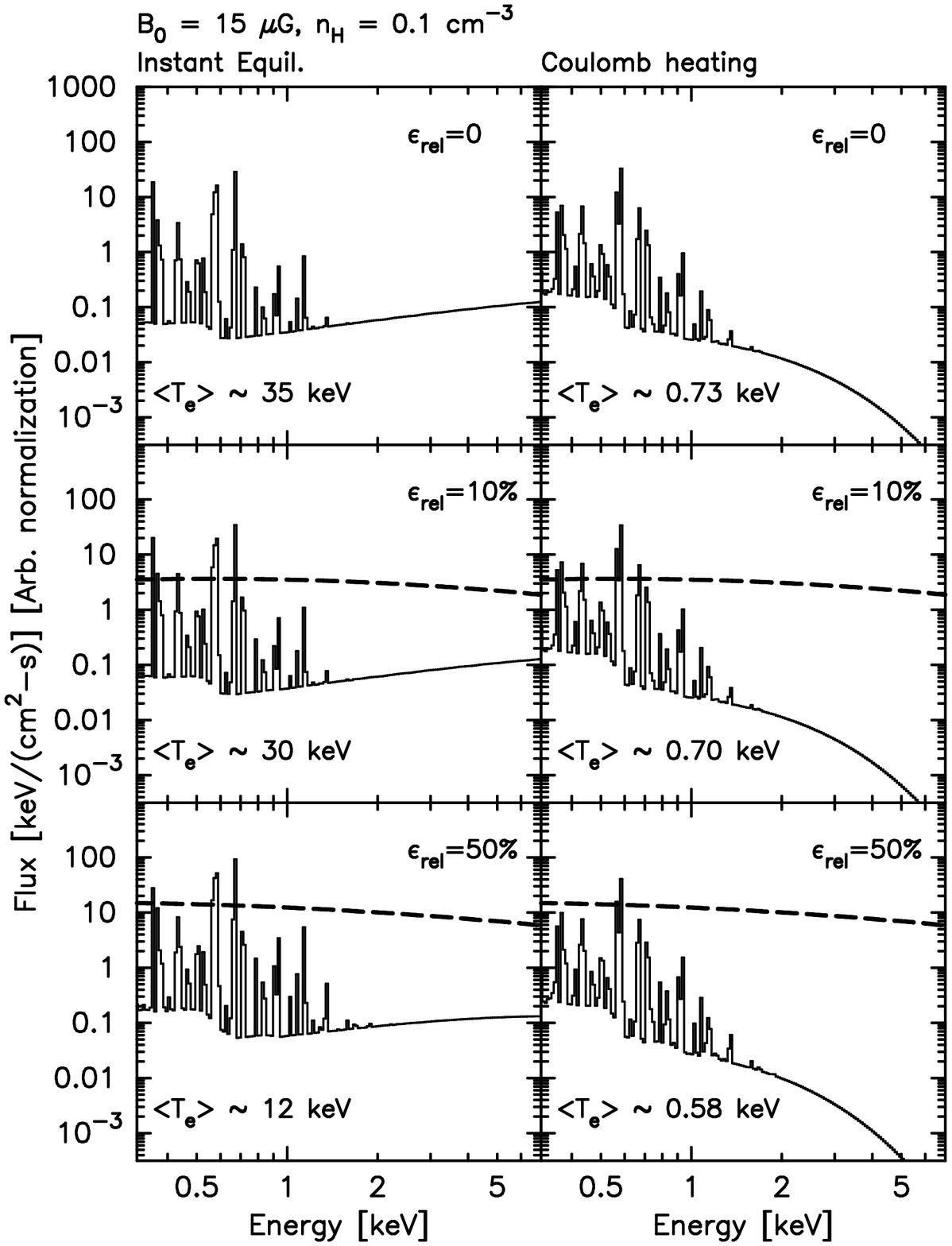} %%6stack_nH0.1_B15_TP_10_50pc_D.eps} 
\caption{Thermal emission for examples where the efficiency of DSA at
  the FS is varied as indicated. The top panels are model A1, the middle
  panels are model A2, and the bottom panels are model A3 in Table~1.
The emission is summed for the region
  between the CD and FS and emission from the RS is not included. The
  dashed curve in each panel is the \syn\ emission from electrons
  accelerated at the FS. Panels on the left assume \IE\ between
  electrons and protons and panels on the right assume \CH\ of
  electrons. All parameters for the various models are identical other
  than $\EffRel$ and the electron heating.
\label{line_eff}}
\end{figure}

\begin{figure}        % Fig 9
\epsscale{.85} \plotone{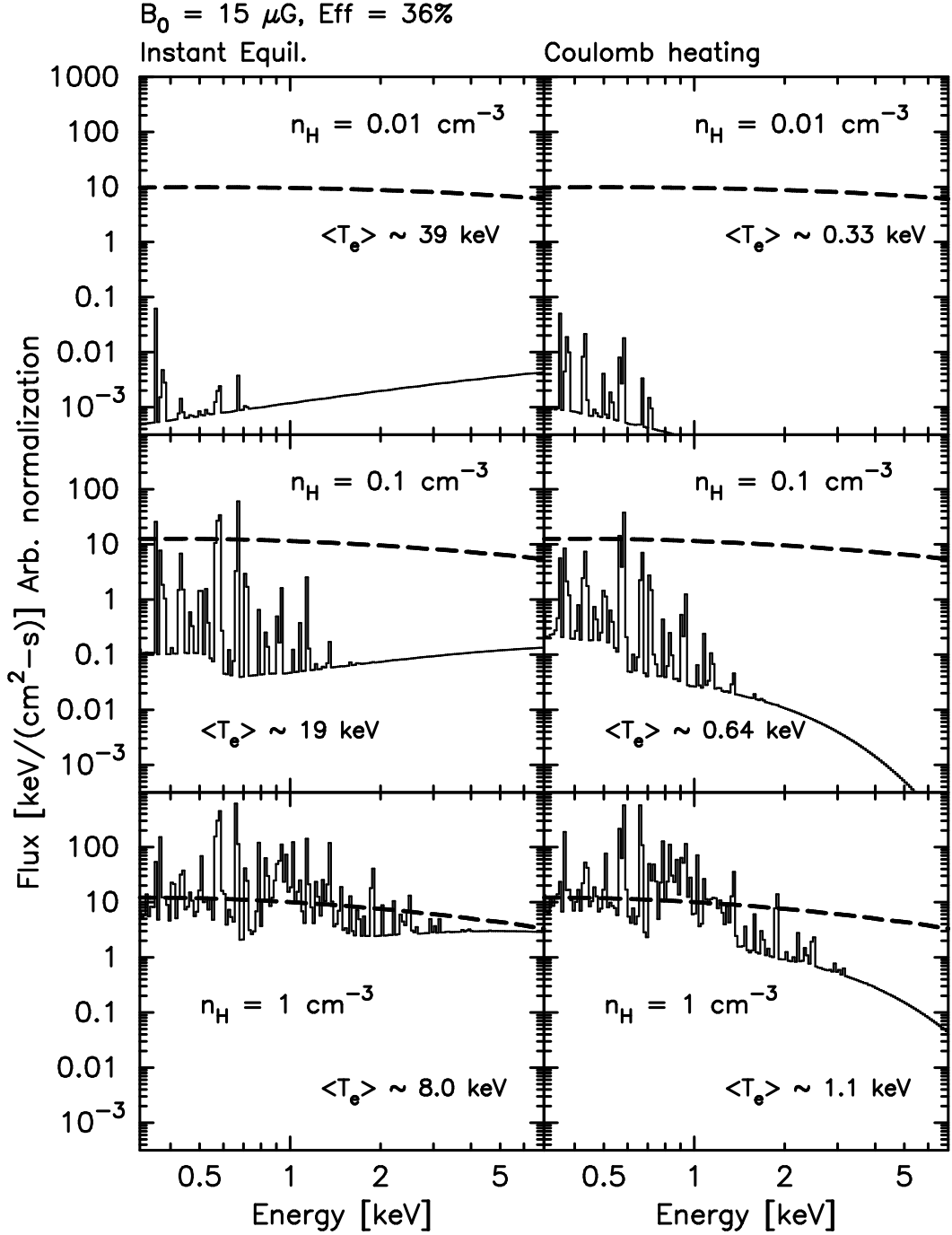} %%6stack_B15_nH_0.01_0.1_1_36pc_D.eps} 
\caption{Examples similar to Figure~\ref{line_eff} where $\EffRel$ is
  set at 36\% and $\denH$ is varied as indicated. The top panels are
  model B1, the middle panels are model B2, and the bottom panels are
  model B3 in Table~1.
\label{line_density}}
\end{figure}

\begin{figure}        % Fig 10
\epsscale{.85} \plotone{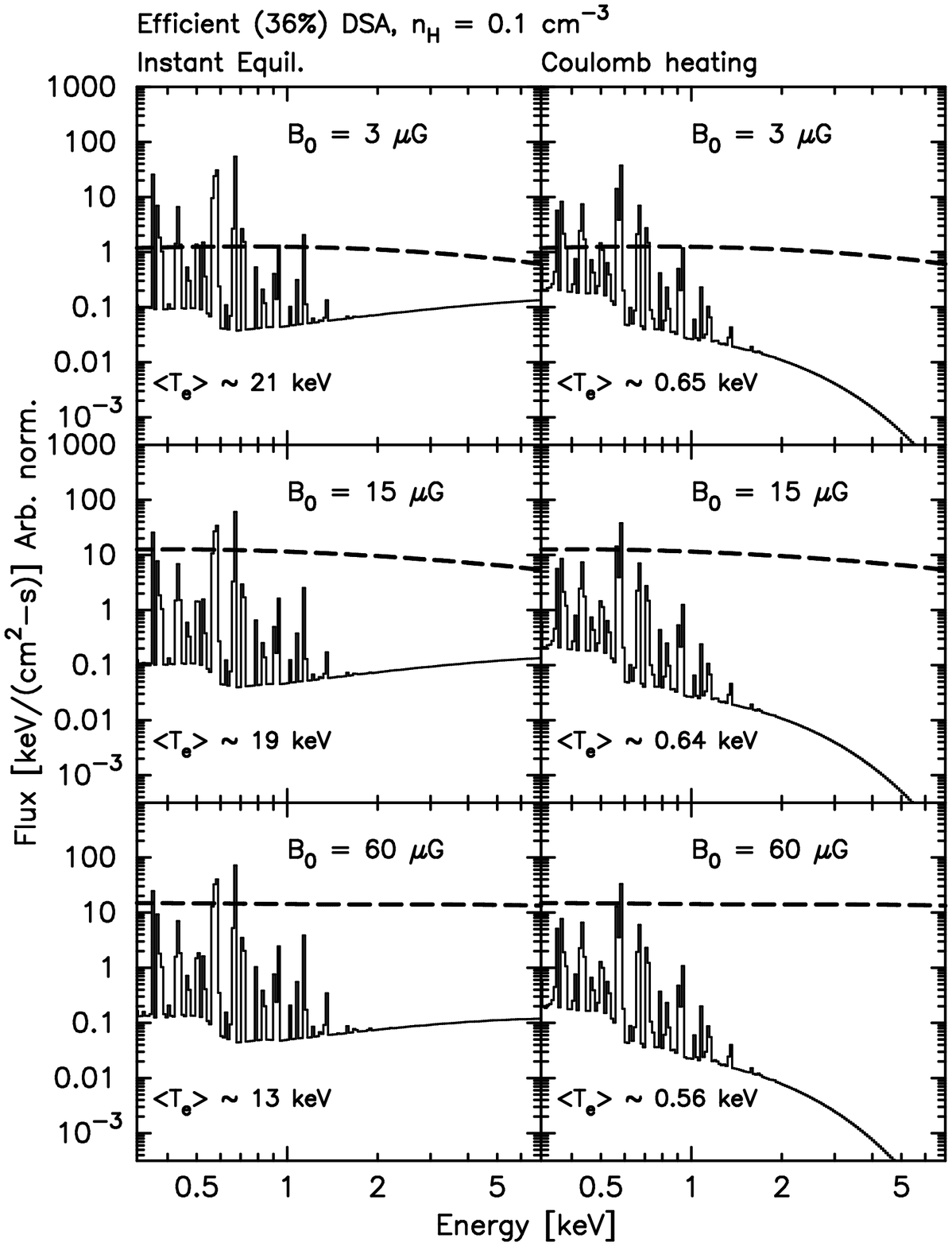} %%6stack_nH0.1_B3_B15_B60_36pc_D.eps} 
\caption{Examples similar to Figure~\ref{line_eff} where $\EffRel$ is
  set at 36\%, $\denH=0.1$\,\pcc\ and the unshocked magnetic field  is
  varied as indicated. The top panels are model C1, the middle
  panels are model C2, and the bottom panels are model C3 in Table~1.
\label{line_B}}
\end{figure}

In the top panel of Figure~\ref{line_Coul} we show a \TP\ example where
no significant population of \rel\ particles is produced. The middle and
bottom panels show the result for efficient DSA, where $\Inj$ has been
chosen so approximately
50\% of the shock energy flux goes into \rel\ particles, mainly
protons.\footnote{In all of our examples, the efficiency is calculated
at the end of the simulation. During earlier times, the instantaneous
acceleration efficiency can be greater or less than the stated value.}
The dashed curve in the middle panel is the \syn\ emission produced by
TeV electrons. Except for a few strong lines,
this dominates over the thermal emission. 
The bottom panel shows the summed thermal and nonthermal spectra from the
middle panel. A direct
comparison with the top panel shows the difference that
results by changing a single parameter, the assumed efficiency for
DSA. For this example, the shocked electrons were heated via Coulomb
collisions with the hot, shocked protons.

In Figures~\ref{line_eff}--\ref{line_B} we show results for models A, B,
and C in Table~1.  We present the results for the two extremes for
electron heating (\IE\ and \CH) to stress the obvious differences in the
thermal continuum which arise from assuming high electron temperatures.
We also vary the acceleration efficiency, $\EffRel$, the CSM density,
$\denH$, and the magnetic field, $B_0$.  In Figure~\ref{line_eff}, we
have varied $\EffRel$ between 0 and 50\% while keeping $B_0=15$\,\muG\
and $\denH=0.1$\,\pcc.  Figure~\ref{line_density} shows the effects of
varying the density $\denH$ while holding $\EffRel$ and $B_0$
constant. Finally, in Figure~\ref{line_B} we vary the magnetic field
while holding $\EffRel$ and $\denH$ constant.  In these figures we plot
the nonthermal synchrotron emission as a dashed line and the thermal
spectrum as a solid line but do not plot the summed spectrum.

We chose to focus mainly on varying the ambient density, acceleration
efficiency, and magnetic field because these parameters have a large
influence on the SNR evolution.  Increasing the upstream density results
in higher volume emission measures and thus higher thermal
continua. Increasing the acceleration efficiency changes the equation of
state of the shocked CSM resulting in lower post shock plasma
temperatures and higher densities. Finally, varying the upstream
magnetic field strength influences the acceleration process because it
is assumed in our model that energy in shock-accelerated particles in
the precursor can be transferred into magnetic turbulence and then into
heat. This heating of the precursor weakens the subshock and results in
less efficient particle acceleration \citep[see][ for a full
discussion]{BE99}. For the cases shown in Figure~\ref{line_B}, the
injection efficiency, $\Inj$ has been adjusted to produce $\EffRel
\simeq 36\%$ as $B_0$ is changed.

\newlistroman

General results from Figures~\ref{line_eff}, \ref{line_density}, and
\ref{line_B} include:

\noindent \listromanDE
\syn\ dominates thermal emission except for the TP examples (top panels
in Figure~\ref{line_eff}) and the $\denH=1$\,\pcc\ examples in the lower
panels in Figure~\ref{line_density}.

\noindent \listromanDE
In the TP case (top panels in Figure~\ref{line_eff}), the differences
between \IE\ and \CH\ are large, with the average electron temperature for
\IE\ $\sim 35$\,keV and only $\sim 0.73$\,keV for \CH.
However, once \syn\ becomes important with $\EffRel \gtrsim 10\%$, the
signature of electron heating is difficult to discern against the \syn\
continuum except for high $\denH$ (i.e., the $\denH=1$\,\pcc\ cases in
Figure~\ref{line_density}).

\noindent \listromanDE 
The cases involving instantaneous equilibration produce models
with anomalously high electron temperatures, even when DSA is efficient. 
While these simulations cannot rule out instantaneous electron heating 
as we are not comparing them to observations of SNR forward shocks, fits
to X-ray spectra do not generally predict such high electron temperatures
in these regions. 

\begin{figure}        % Fig 11
\epsscale{.60} \plotone{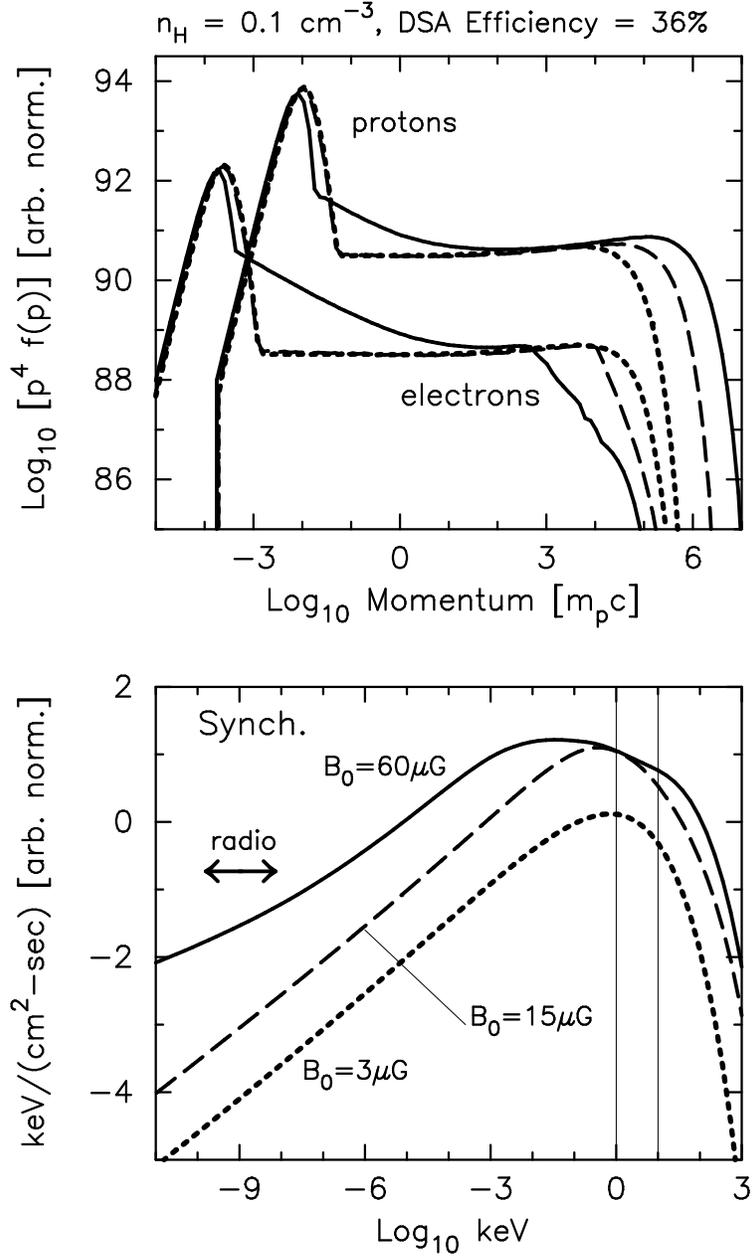} %%fp_syn_NL_n0.1_B3_B15_B60_36pc_40.eps} 
\caption{The top panel shows particle spectra summed over the SNR at the
end of the simulation, i.e., $\tSNR=500$\,yr. The bottom panel shows the
\syn\ emission from these electrons. All models have the input same
parameters except for $B_0$ which is 3\,\muG\ for the dotted curves
(model C1), 15\,\muG\ for the dashed curves (model C2), and 60\,\muG\
for the solid curves (model C3).
\label{fp_syn}}
\end{figure}

\begin{figure}        % Fig 12
\epsscale{.60} \plotone{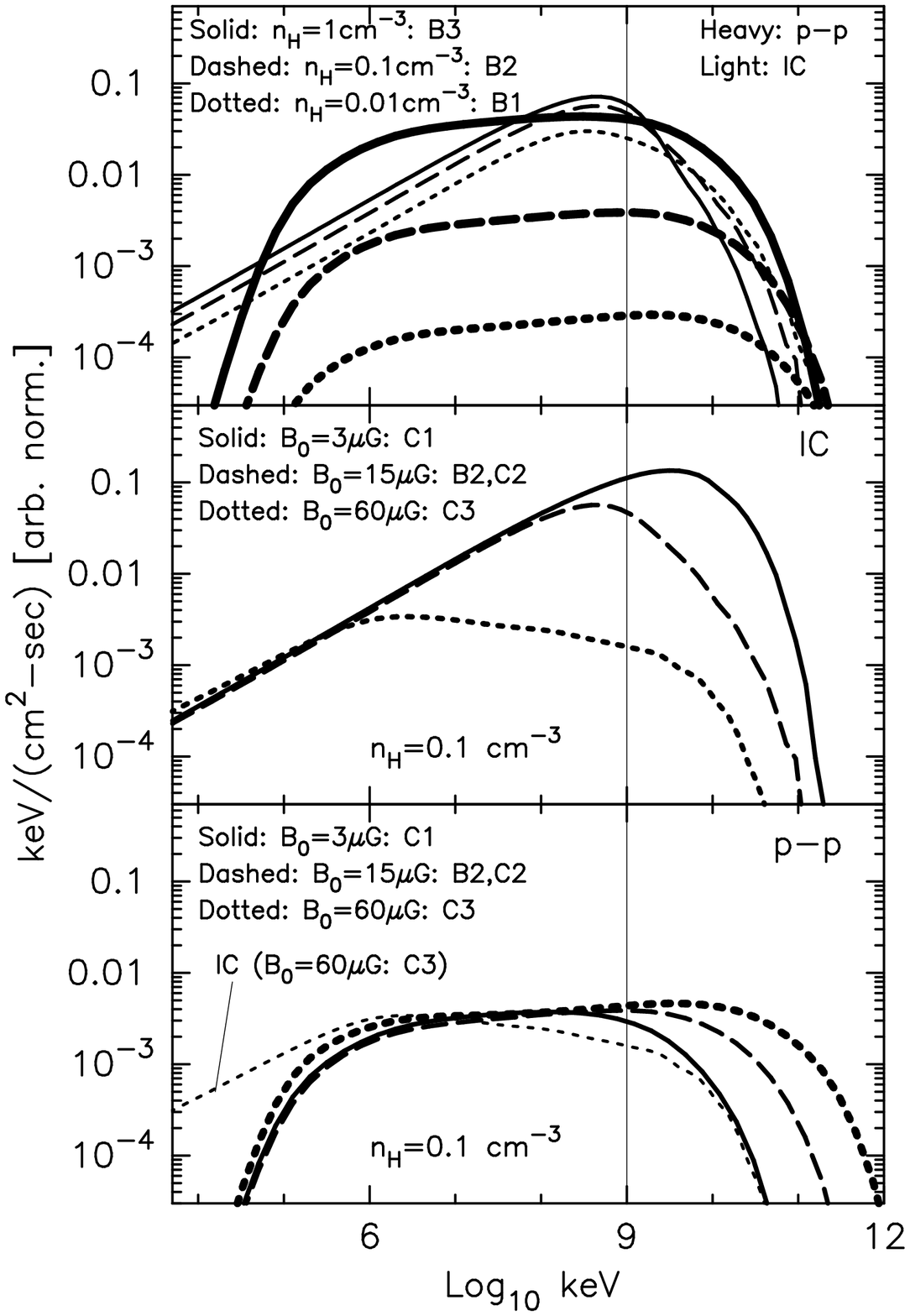} %%pion_IC_3stack_40_b.eps} 
\caption{Pion-decay and \IC\ emission for a range of $\denH$ and
  $B_0$. In the top panel, the heavy-weight curves are \pion, the
  light-weight curves are \IC, and $\EffRel=36\%$ and $B_0=15$\,\muG\ in
  all cases. The strong dependence of \pion\ on ambient density $\denH$
  is evident.  The middle panel shows IC and the bottom panel shows
  \pion\ for $\denH=0.1$\,\pcc\ with $B_0$ varying from 3\,\muG\ (solid
  curves) to 15\,\muG\ (dashed curves) to 60\,\muG\ (dotted curves). For
  comparison to the \piZ, we show in the bottom panel the IC emission
  for $B_0=60$\,\muG\ (light-weight dotted curve).
The particle distributions producing the emission in the bottom two
  panels are those shown in the top panel of Figure~\ref{fp_syn}.
\label{IC_pion}}
\end{figure}

In the top panel of Figure~\ref{fp_syn} we show the electron and proton
spectra for the three \IE\ examples of Figure~\ref{line_B}. The bottom
panel shows the \syn\ spectra produced by these electrons.
The three cases have $\EffRel \simeq 36\%$ even though the proton
spectra look quite different. The major reason for this is that
$\EffRel$ includes the energy in particles that escape upstream from the
shock and which is not included in the plotted spectra.
The spectra shown are summed over the interaction region and the
acceleration efficiency varies during the evolution of the remnant.

As $B_0$ increases, the maximum proton momentum increases in approximate
proportion to $B_0$.\footnote{We note that this simple scaling is
almost certain not to hold if magnetic field amplification is
considered. A full discussion is beyond the scope of this paper but
preliminary work \citep*{AB2006,VEB2006,BAC2006} suggests that $\pmax$
will not increase in proportion to $B_0$.}
The maximum electron energy
decreases with $B_0$, however, because \syn\ losses become more severe. 
The break in the electron spectrum for the $B_0=60$\,\muG\ case near
$10^3 m_pc$ (solid curves) results because strong radiation losses lower
$\Pmax$ for shells shocked early-on. When all shells are added together
at $\tSNR$, a steep region spanning several decades in momentum results
and this produces a relatively flat region in the X-ray \syn\ spectrum
shown in the bottom panel.
In the 1--10 keV X-ray energy range, the \syn\ flux varies by nearly an
order of magnitude between $B_0=3$ and 15\,\muG, but the shape is almost
indistinguishable.

The GeV-TeV electrons produce \IC\ emission as well as \syn\ radiation,
and the protons accelerated by the FS interact and produce GeV-TeV
photons via proton-proton collisions and \pion.
In the top panel of Figure~\ref{IC_pion} we compare the \IC\ and \pion\
emission for the examples shown in Figure~\ref{line_density} where
$\denH$ varies and $B=15$\,\muG. As expected, the overall \pion\
flux increases more rapidly with $\denH$ than the \IC\ flux. At 1 TeV,
the ratio \pionIC\ is 0.01 for $\denH=0.01$\,\pcc, 0.08 for
$\denH=0.1$\,\pcc, and 0.65 for $\denH=1$\,\pcc.  It is noteworthy that
in order to obtain (pion/IC$)_\mathrm{1TeV} > 1$ densities greater than
1\,\pcc\ are required.
At energies below $\sim 100$\,GeV, the \pion\ emission for
$\denH=1$\,\pcc\ becomes stronger than the \IC\ as it does at energies
above a few TeV.
The energy range that will be explored by the Large Area Telescope (LAT)
on GLAST (20\,MeV--300\,GeV) should allow a clear differentiation
between emission processes in this case.

In the bottom two panels of Figure~\ref{IC_pion} we show the IC and
pion-decay emission for the particle spectra shown in the top panel of
Figure~\ref{fp_syn}.  The radiation losses caused by the large
magnetic field (dotted curves: $B_0=60$\,\muG\ and $B_2=310$\,\muG),
combined with the remnant evolution, produce a break in the IC emission
in the GeV--TeV range.
A comparison of the light- and heavy-weight dotted curves in the bottom
panel shows that the shapes of the IC and \piZ\ components are less
distinct than with lower fields and this may make it more difficult to
distinguish these components.  When $B_0$ is large, however, the \pion\
emission extends to much higher energy and combining the observations of
GLAST and ground-based air-Cherenkov telescopes will help in
distinguishing between the hadronic and leptonic scenarios.  Currently,
HESS spectra of SNRs roughly span the energy range $0.1-10$ TeV and
above 1 TeV the difference between the two dotted curves in the bottom
panel becomes significant.

\section{DISCUSSION AND CONCLUSIONS}
The X-ray emission from SNR shocks is generally a mixture of thermal
continuum and line emission, and nonthermal continuum. These components
are connected through the particle acceleration process, but are
generally modeled independently. Here we present the first results from
an ongoing project to model the X-ray emission in a fully \SC\ manner
with the SNR hydrodynamics and broad-band continuum radiation.  The
problem is a complicated one with a number of uncertain parameters both
for the model and the environment. In order to simplify our presentation
and to emphasize the important aspects of our model, we
%only calculate 
concentrate on 
emission from the outer blast wave (forward
shock) and use approximate techniques to model the \NEI\ in the
interaction region between the \CD\ and FS. 

The results we present are intended to show broadly how the X-ray
emission depends on important parameters including the shock
acceleration efficiency, $\EffRel$, the unshocked CSM proton number
density, $\denH$, the magnetic field strength, $B_0$, and two extremes
of electron heating: \IE\ and \CH.
\newlistroman
In summary, our results for the forward shock show:

\noindent\listromanDE \Syn\ emission from \rel\ electrons becomes
important, and even dominant, in the X-ray energy range for modest
acceleration efficiencies over a wide range of CSM densities and magnetic
fields. As shown in Figures~\ref{line_eff}, \ref{line_density}, and
\ref{line_B}, only TP cases or those with large $\denH\sim 1$\,\pcc\
show thermal line emission clearly dominant over \syn\ emission. Once $\EffRel
\gtrsim 10\%$ \syn\ emission is likely to be important for typical SNR
parameters. 

\noindent\listromanDE The two extremes for electron heating, \IE\ and
\CH, produce very different thermal spectra as expected. However, it may
be difficult to distinguish these in cases where \syn\ radiation dominates.

\noindent\listromanDE Changes in the ambient magnetic field, $B_0$, have
a large effect on the \syn\ emission in the X-ray range but these
effects tend to saturate in terms of overall flux for large $B_0$
(bottom four panels in Figure~\ref{line_B}).  The saturation is a result
of the fact that large magnetic fields tend to dampen DSA, at least in
the approximate DSA model used here.

\noindent\listromanDE As has been recognized for some time, broad-band
models and fits are critical for constraining important SNR parameters;
we illustrate this in Figures~\ref{fp_syn} and \ref{IC_pion}. In
homogeneous \NL\ models, such as ours, changes in parameters that modify
emission in one energy band will modify emission from radio to TeV
\gamrays. Thus, if large magnetic fields produce hard \syn\ spectra in
the X-ray band, these same electrons will produce strong radio emission
(bottom panel of Figure~\ref{fp_syn}). If particular ambient densities
are required to match thermal X-ray observations, these densities
strongly influence the \pionIC\ ratio (top panel in
Figure~\ref{IC_pion}). 

\noindent\listromanDE As has been emphasized by many authors, direct
evidence for CR ion production in SNRs can come from identifying \pion\
emission at GeV--TeV energies. 
The interpretation of GeV--TeV observations is complicated however
because \pion\  photons from proton-proton
interactions and IC emission from TeV electrons scattering off of the
cosmic background radiation are expected to produce similar fluxes in
the GeV-TeV energy range.
It is generally assumed that the shape of \pion\ emission is
sufficiently different from that of IC to allow a clear differentiation
if a large enough energy range is sampled.  In the bottom two panels of
Figure~\ref{IC_pion} we show spectra for a range of $B_0$ and it is
clear that the shapes of the IC spectra in the GeV-TeV region depend
fairly strongly on the magnetic field with the $B_0=60$\,\muG\ case
(where strong radiation losses and evolution produce a break in the
underlying electron spectrum) closely mimicking the \pion\ shape in the
1 GeV--1 TeV range.
For large fields, however, significant differences in shape and
intensity occur at both lower and higher energies highlighting the
importance of air-Cherenkov observations above 1~TeV and for 
furure observations by GLAST below 1 GeV.

\newlistroman The model we have presented is, we believe, the first to
\SCly\ determine the \syn\ continuum with thermal emission in the X-ray
energy range. However, a number of steps remain before realistic models
are produced that can be used to interpret X-ray observations.  
These steps include:
\listromanDE modeling of the reverse shock, which requires an accurate
description of the ejecta composition and spatial distribution;
\listromanDE calculating the ionization rates for material behind both
the forward and reverse shocks, including time dependence and variations
in temperature and electron density;
\listromanDE line-of-sight projections; and 
\listromanDE incorporating the effects of the instrument response for
general X-ray spectra without restriction to specific models such as
those currently used in, for example, XSPEC.

\acknowledgments
D.C.E is grateful for support from a NASA ATP grant (ATP02-0042-0006)
and a NASA LTSA grant (NNH04Zss001N-LTSA), P.O.S. acknowledges support
from NASA contract NAS8-39073, and the work of P.B. was partially funded
through grant Prin2004.

\newpage

% bbbb  Note: must have files:  aa.bst  and  aa.cls
\bibliographystyle{aa} % A&A style
%\bibliography{c:/a_a_TOP/a_active/bibTeX/bib_DCE}
\bibliography{bib_DCE}

\begin{thebibliography}{47}
\expandafter\ifx\csname natexlab\endcsname\relax\def\natexlab#1{#1}\fi

\bibitem[{{Aharonian} {et~al.}(2004){Aharonian}, {Akhperjanian}, {Aye},
  {Bazer-Bachi}, {Beilicke}, {Benbow}, {Berge}, {Berghaus}, {Bernl{\" o}hr},
  {Bolz}, {Boisson}, {Borgmeier}, {Breitling}, {Brown}, {Bussons Gordo},
  {Chadwick}, {Chitnis}, {Chounet}, {Cornils}, {Costamante}, {Degrange},
  {Djannati-Ata{\" i}}, {Drury}, {Ergin}, {Espigat}, {Feinstein}, {Fleury},
  {Fontaine}, {Funk}, {Gallant}, {Giebels}, {Gillessen}, {Goret}, {Guy},
  {Hadjichristidis}, {Hauser}, {Heinzelmann}, {Henri}, {Hermann}, {Hinton},
  {Hofmann}, {Holleran}, {Horns}, {de Jager}, {Jung}, {Kh{\' e}lifi}, {Komin},
  {Konopelko}, {Latham}, {Le Gallou}, {Lemoine}, {Lemi{\` e}re}, {Leroy},
  {Lohse}, {Marcowith}, {Masterson}, {McComb}, {de Naurois}, {Nolan},
  {Noutsos}, {Orford}, {Osborne}, {Ouchrif}, {Panter}, {Pelletier}, {Pita},
  {Pohl}, {P{\" u}hlhofer}, {Punch}, {Raubenheimer}, {Raue}, {Raux}, {Rayner},
  {Redondo}, {Reimer}, {Reimer}, {Ripken}, {Rivoal}, {Rob}, {Rolland},
  {Rowell}, {Sahakian}, {Saug{\' e}}, {Schlenker}, {Schlickeiser}, {Schuster},
  {Schwanke}, {Siewert}, {Sol}, {Steenkamp}, {Stegmann}, {Tavernet}, {Th{\'
  e}oret}, {Tluczykont}, {van der Walt}, {Vasileiadis}, {Vincent}, {Visser},
  {V{\" o}lk}, \& {Wagner}}]{AharonianNature2004}
{Aharonian}, F.~A., {Akhperjanian}, A.~G., {Aye}, K.-M., {et~al.} 2004, \nat,
  432, 75

\bibitem[{{Albert} {et~al.}(2006){Albert}, {Aliu}, {Anderhub}, {Antoranz},
  {Armada}, {Asensio}, {Baixeras}, {Barrio}, {Bartel}, {Bartko}, {Bastieri},
  {Bavikadi}, {Bednarek}, {Berger}, {Bigongiari}, {Biland}, {Bisesi}, {Blanch},
  {Bock}, {Bretz}, {Britvitch}, {Camara}, {Chilingarian}, {Ciprini}, {Coarasa},
  {Commichau}, {Contreras}, {Cortina}, {Curtev}, {Danielyan}, {Dazzi}, {De
  Angelis}, {de los Reyes}, {De Lotto}, {Domingo-Santamaria}, {Dorner}, {Doro},
  {Errando}, {Fagiolini}, {Ferenc}, {Fern{\'a}ndez}, {Firpo}, {Flix},
  {Fonseca}, {Font}, {Galante}, {Garczarczyk}, {Gaug}, {Gebauer}, {Giller},
  {Goebel}, {Hakobyan}, {Hayashida}, {Hengstebeck}, {H{\"o}hne}, {Hose},
  {Jacon}, {Kalekin}, {Kranich}, {Laille}, {Lenisa}, {Liebing}, {Lindfors},
  {Longo}, {L{\'o}pez}, {L{\'o}pez}, {Lorenz}, {Lucarelli}, {Majumdar},
  {Maneva}, {Mannheim}, {Mariotti}, {Mart{\'{\i}}nez}, {Mase}, {Mazin},
  {Merck}, {Merck}, {Meucci}, {Meyer}, {Miranda}, {Mirzoyan}, {Mizobuchi},
  {Moralejo}, {Nilsson}, {O{\~n}a-Wilhelmi}, {Ordu{\~n}a}, {Otte}, {Oya},
  {Paneque}, {Paoletti}, {Pasanen}, {Pascoli}, {Pauss}, {Pavel}, {Pegna},
  {Peruzzo}, {Piccioli}, {Prandini}, {Rico}, {Rhode}, {Riegel}, {Rissi},
  {Robert}, {Rossato}, {R{\"u}gamer}, {Saggion}, {Sanchez}, {Sartori},
  {Scalzotto}, {Schmitt}, {Schweizer}, {Shayduk}, {Shinozaki}, {Shore},
  {Sidro}, {Sillanp{\"a}{\"a}}, {Sobczynska}, {Stamerra}, {Stark}, {Takalo},
  {Temnikov}, {Tescaro}, {Teshima}, {Tonello}, {Torres}, {Torres}, {Turini},
  {Vankov}, {Vitale}, {Wagner}, {Wibig}, {Wittek}, \&
  {Zapatero}}]{MAGIC_J1813_2006}
{Albert}, J., {Aliu}, E., {Anderhub}, H., {et~al.} 2006, \apjl, 637, L41

\bibitem[{{Allen} {et~al.}(2005){Allen}, {Houck}, \& {Sturner}}]{AllenEtal2005}
{Allen}, G.~E., {Houck}, J.~C., \& {Sturner}, S.~J. 2005, in X-Ray and Radio
  Connections (eds. L.O. Sjouwerman and K.K Dyer) Published electronically by
  NRAO, http://www.aoc.nrao.edu/events/xraydio Held 3-6 February 2004 in Santa
  Fe, New Mexico, USA, (E4.07) 6 pages

\bibitem[{{Amato} \& {Blasi}(2005)}]{AB2005}
{Amato}, E. \& {Blasi}, P. 2005, \mnras, 364, L76

\bibitem[{{Amato} \& {Blasi}(2006)}]{AB2006}
{Amato}, E. \& {Blasi}, P. 2006, \mnras, 371, 1251

\bibitem[{Baring {et~al.}(1999)Baring, Ellison, Reynolds, Grenier, \&
  Goret}]{BaringEtal99}
Baring, M.~G., Ellison, D.~C., Reynolds, S.~P., Grenier, I.~A., \& Goret, P.
  1999, ApJ, 513, 311

\bibitem[{Bell \& Lucek(2001)}]{BL2001}
Bell, A.~R. \& Lucek, S.~G. 2001, MNRAS, 321, 433

\bibitem[{Berezhko \& Ellison(1999)}]{BE99}
Berezhko, E.~G. \& Ellison, D.~C. 1999, ApJ, 526, 385

\bibitem[{{Berezhko} {et~al.}(2002){Berezhko}, {Ksenofontov}, \& {V{\"
  o}lk}}]{BKV2002}
{Berezhko}, E.~G., {Ksenofontov}, L.~T., \& {V{\" o}lk}, H.~J. 2002, \aap, 395,
  943

\bibitem[{{Blasi}(2002)}]{Blasi2002}
{Blasi}, P. 2002, Astroparticle Physics, 16, 429

\bibitem[{{Blasi}(2004)}]{Blasi2004}
{Blasi}, P. 2004, Astroparticle Physics, 21, 45

\bibitem[{{Blasi} {et~al.}(2006){Blasi}, {Amato}, \& {Caprioli}}]{BAC2006}
{Blasi}, P., {Amato}, E., \& {Caprioli}, D. 2006, ArXiv Astrophysics e-prints

\bibitem[{{Blasi} {et~al.}(2005){Blasi}, {Gabici}, \& {Vannoni}}]{BGV2005}
{Blasi}, P., {Gabici}, S., \& {Vannoni}, G. 2005, \mnras, 361, 907

\bibitem[{{Borkowski} {et~al.}(2001){Borkowski}, {Lyerly}, \&
  {Reynolds}}]{borkowski01}
{Borkowski}, K.~J., {Lyerly}, W.~J., \& {Reynolds}, S.~P. 2001, \apj, 548, 820

\bibitem[{{Boulares} \& {Cox}(1988)}]{BoularesCox88}
{Boulares}, A. \& {Cox}, D.~P. 1988, \apj, 333, 198

\bibitem[{Chevalier(1983)}]{ch83}
Chevalier, R.~A. 1983, ApJ, 272, 765

\bibitem[{Decourchelle {et~al.}(2000)Decourchelle, Ellison, \&
  Ballet}]{DEB2000}
Decourchelle, A., Ellison, D.~C., \& Ballet, J. 2000, ApJ, 543, L57

\bibitem[{{DeLaney} {et~al.}(2002){DeLaney}, {Koralesky}, {Rudnick}, \&
  {Dickel}}]{DeLaneyEtal2002}
{DeLaney}, T., {Koralesky}, B., {Rudnick}, L., \& {Dickel}, J.~R. 2002, \apj,
  580, 914

\bibitem[{{Dorfi} \& {Bohringer}(1993)}]{DorfiBohr93}
{Dorfi}, E.~A. \& {Bohringer}, H. 1993, \aap, 273, 251

\bibitem[{{Dwarkadas}(2000)}]{Dwarkadas2000}
{Dwarkadas}, V.~V. 2000, \apj, 541, 418

\bibitem[{{Eichler}(1984)}]{Eichler84}
{Eichler}, D. 1984, \apj, 277, 429

\bibitem[{Ellison {et~al.}(2000)Ellison, Berezhko, \& Baring}]{EBB2000}
Ellison, D.~C., Berezhko, E.~G., \& Baring, M.~G. 2000, ApJ, 540, 292

\bibitem[{{Ellison} \& {Cassam-Chena{\"{\i}}}(2005)}]{EC2005}
{Ellison}, D.~C. \& {Cassam-Chena{\"{\i}}}, G. 2005, \apj, 632, 920

\bibitem[{Ellison {et~al.}(2004)Ellison, Decourchelle, \& Ballet}]{EDB2004}
Ellison, D.~C., Decourchelle, A., \& Ballet, J. 2004, A\&A, 413, 189

\bibitem[{{Ellison} {et~al.}(2005){Ellison}, {Decourchelle}, \&
  {Ballet}}]{EDB2005}
{Ellison}, D.~C., {Decourchelle}, A., \& {Ballet}, J. 2005, \aap, 429, 569

\bibitem[{{Ellison} {et~al.}(2001){Ellison}, {Slane}, \& {Gaensler}}]{ESG2001}
{Ellison}, D.~C., {Slane}, P., \& {Gaensler}, B.~M. 2001, \apj, 563, 191

\bibitem[{{Gotthelf} {et~al.}(2001){Gotthelf}, {Koralesky}, {Rudnick}, {Jones},
  {Hwang}, \& {Petre}}]{GotthelfEtal2001}
{Gotthelf}, E.~V., {Koralesky}, B., {Rudnick}, L., {et~al.} 2001, \apjl, 552,
  L39

\bibitem[{{Hamilton} {et~al.}(1983){Hamilton}, {Chevalier}, \&
  {Sarazin}}]{HSC83}
{Hamilton}, A.~J.~S., {Chevalier}, R.~A., \& {Sarazin}, C.~L. 1983, \apjs, 51,
  115

\bibitem[{{Hamilton} \& {Sarazin}(1984)}]{hamilton84}
{Hamilton}, A.~J.~S. \& {Sarazin}, C.~L. 1984, \apj, 284, 601

\bibitem[{{Heavens}(1984)}]{Heavens84}
{Heavens}, A.~F. 1984, \mnras, 211, 195

\bibitem[{{Hughes} \& {Helfand}(1985)}]{hughes85}
{Hughes}, J.~P. \& {Helfand}, D.~J. 1985, \apj, 291, 544

\bibitem[{{Itoh}(1979)}]{itoh79}
{Itoh}, H. 1979, \pasj, 31, 541

\bibitem[{{Jones} \& {Ellison}(1991)}]{JE91}
{Jones}, F.~C. \& {Ellison}, D.~C. 1991, Space Science Reviews, 58, 259

\bibitem[{{Jones} {et~al.}(2003){Jones}, {Rudnick}, {DeLaney}, \&
  {Bowden}}]{JonesEtal2003}
{Jones}, T.~J., {Rudnick}, L., {DeLaney}, T., \& {Bowden}, J. 2003, \apj, 587,
  227

\bibitem[{{LeBohec} {et~al.}(2006){LeBohec}, {Atkins}, {Badran}, {Blaylock},
  {Bond}, {Boyle}, {Bradbury}, {Buckley}, {Carter-Lewis}, {Celik}, {Chow},
  {Cogan}, {P}, {Daniel}, {de la Calle Perez}, {Dowdall}, {Dowkontt}, {Duke},
  {Ergin}, {Falcone}, {Fegan}, {Fegan}, {Finley}, {Fortin}, {Fortson},
  {Gammell}, {Gibbs}, {Gillanders}, {Grube}, {Hall}, {Hanna}, {Hays}, {Holder},
  {Horana}, {Hughes}, {Humensky}, {Kaaret}, {Kenny}, {Kertzmann}, {Kieda},
  {Kildea}, {Knapp}, {Kosack}, {Krawczynski}, {Krennrich}, {Lang}, {Linton},
  {Lloyd-Evans}, {Maier}, {Manseri}, {Milovanovic}, {Moriarty}, {Mukherjee},
  {Nagai}, {Ogden}, {Olevitch}, {Ong}, {Perkins}, {Petry}, {Pizlo}, {Pohl},
  {Power-Mooney}, {Quinn}, {Quinn}, {Ragan}, {Reynolds}, {Rebillot}, {Rose},
  {Schroedter}, {Sembroski}, {Steele}, {Swordy}, {Syson}, {Valcarcel},
  {Vassiliev}, {Wagner}, {Wakely}, {Walker}, {Weekes}, {White}, {Williams}, \&
  {Zweerink}}]{LeBohec2006}
{LeBohec}, S., {Atkins}, R.~W., {Badran}, H.~M., {et~al.} 2006, Journal of
  Physics Conference Series, 47, 232

\bibitem[{{Porquet} {et~al.}(2001){Porquet}, {Arnaud}, \&
  {Decourchelle}}]{Porquet2001}
{Porquet}, D., {Arnaud}, M., \& {Decourchelle}, A. 2001, \aap, 373, 1110

\bibitem[{{Reynolds} \& {Ellison}(1992)}]{RE92}
{Reynolds}, S.~P. \& {Ellison}, D.~C. 1992, \apjl, 399, L75

\bibitem[{{Rho} {et~al.}(2002){Rho}, {Dyer}, {Borkowski}, \&
  {Reynolds}}]{RhoEtal2002}
{Rho}, J., {Dyer}, K.~K., {Borkowski}, K.~J., \& {Reynolds}, S.~P. 2002, \apj,
  581, 1116

\bibitem[{{Slane} {et~al.}(1999){Slane}, {Gaensler}, {Dame}, {Hughes},
  {Plucinsky}, \& {Green}}]{Slane99}
{Slane}, P., {Gaensler}, B.~M., {Dame}, T.~M., {et~al.} 1999, \apj, 525, 357

\bibitem[{{Spitzer}(1968)}]{spitzer68}
{Spitzer}, L. 1968, {Diffuse matter in space} (New York: Interscience
  Publication, 1968)

\bibitem[{{Tanimori} {et~al.}(1998){Tanimori}, {Hayami}, {Kamei}, {Dazeley},
  {Edwards}, {Gunji}, {Hara}, {Hara}, {Holder}, {Kawachi}, {Kifune}, {Kita},
  {Konishi}, {Masaike}, {Matsubara}, {Matsuoka}, {Mizumoto}, {Mori}, {Moriya},
  {Muraishi}, {Muraki}, {Naito}, {Nishijima}, {Oda}, {Ogio}, {Patterson},
  {Roberts}, {Rowell}, {Sakurazawa}, {Sako}, {Sato}, {Susukita}, {Suzuki},
  {Suzuki}, {Tamura}, {Thornton}, {Yanagita}, {Yoshida}, \&
  {Yoshikoshi}}]{TanimoriEtal1998}
{Tanimori}, T., {Hayami}, Y., {Kamei}, S., {et~al.} 1998, \apjl, 497, L25+

\bibitem[{{V{\" o}lk} {et~al.}(2002){V{\" o}lk}, {Berezhko}, {Ksenofontov}, \&
  {Rowell}}]{VBKR2002}
{V{\" o}lk}, H.~J., {Berezhko}, E.~G., {Ksenofontov}, L.~T., \& {Rowell}, G.~P.
  2002, \aap, 396, 649

\bibitem[{{Vink} {et~al.}(1997){Vink}, {Kaastra}, \& {Bleeker}}]{Vink97}
{Vink}, J., {Kaastra}, J.~S., \& {Bleeker}, J.~A.~M. 1997, \aap, 328, 628

\bibitem[{{Vink} \& {Laming}(2003)}]{VL2003}
{Vink}, J. \& {Laming}, J.~M. 2003, \apj, 584, 758

\bibitem[{{Vladimirov} {et~al.}(2006){Vladimirov}, {Ellison}, \&
  {Bykov}}]{VEB2006}
{Vladimirov}, A., {Ellison}, D.~C., \& {Bykov}, A. 2006, \apj, 652, 1246

\bibitem[{{V{\"o}lk} {et~al.}(2005a){V{\"o}lk}, {Berezhko}, \&
  {Ksenofontov}}]{VBK2005a}
{V{\"o}lk}, H.~J., {Berezhko}, E.~G., \& {Ksenofontov}, L.~T. 2005a, \aap, 433,
  229

\bibitem[{{Warren} {et~al.}(2005){Warren}, {Hughes}, {Badenes}, {Ghavamian},
  {McKee}, {Moffett}, {Plucinsky}, {Rakowski}, {Reynoso}, \&
  {Slane}}]{WarrenEtal2005}
{Warren}, J.~S., {Hughes}, J.~P., {Badenes}, C., {et~al.} 2005, \apj, 634, 376

\end{thebibliography}

\end{document}